\documentclass[fleqn,usenatbib]{mnras}

\usepackage{newtxtext,newtxmath}

\usepackage[T1]{fontenc}

\DeclareRobustCommand{\VAN}[3]{#2}
\let\VANthebibliography\thebibliography
\def\thebibliography{\DeclareRobustCommand{\VAN}[3]{##3}\VANthebibliography}

\usepackage{graphicx}	
\usepackage{amsmath}	
\usepackage{enumitem}

\newcommand{\Omegab}{\Omega_\mathrm{b}}
\newcommand{\bOmegab}{\boldsymbol{\Omega}_\mathrm{b}}
\newcommand{\HJ}{H_\mathrm{J}}

\title[Bar-driven dispersal of Galactic substructure]{Bar-driven dispersal of Galactic substructure}

\author[A. M. Dillamore et al.]{
Adam M. Dillamore\thanks{E-mail: a.dillamore@ucl.ac.uk (AMD)}
and Jason L. Sanders
\\
Department of Physics and Astronomy, University College London, London, WC1E 6BT, UK\\
}

\date{Accepted XXX. Received YYY; in original form ZZZ}

\pubyear{\the\year{}}

\begin{document}
\label{firstpage}
\pagerange{\pageref{firstpage}--\pageref{lastpage}}
\maketitle

\begin{abstract}
Galactic archaeologists often assume that integrals of motion (IoMs) such as $L_z$ and $E$ are conserved, so substructure remains frozen in IoM space over many Gyr. However, this is not true in the Milky Way due in part to its rotating bar. In this study we quantify the effects of the bar on the dynamics of substructure. We employ three different theoretical models: an analytical toy model; a set of test particle simulations with steady and slowing bars; and a cosmological zoom-in simulation of a Milky Way-like galaxy. Each model predicts that the bar increases the angular momentum and energy spread of low-energy substructures by a factor of $\sim10-100$, so they cannot remain tightly clustered. We derive a criterion for determining when this effect is important. The most affected orbits are low energy ($E\lesssim E_\odot$, $r_\mathrm{apo}<40$~kpc), prograde, eccentric, or low inclination. This includes $\sim3/4$ of Galactic globular clusters and $\sim1/4$ of known stellar streams. We predict the presence of abundant bar-dispersed substructure. The structures remain much more tightly clustered in the space of metallicity and Jacobi integral $H_\mathrm{J}=E-\Omega_\mathrm{b}L_z$.  We therefore propose using $H_\mathrm{J}$ and chemistry instead of traditional IoMs when searching for inner halo substructure. In $(L_z,E)$ space the dispersal of the structures is along a principal direction with gradient $\mathrm{d}E/\mathrm{d}L_z$ equal to the bar's pattern speed $\Omega_\mathrm{b}$. Bar-dispersed substructure should therefore allow the past evolution of $\Omega_\mathrm{b}$ to be constrained.
\end{abstract}

\begin{keywords}
Galaxy: kinematics and dynamics -- Galaxy: halo -- Galaxy: structure
\end{keywords}



\section{Introduction}
In the Lambda-CDM ($\Lambda$CDM) model of cosmology, galaxies form hierarchically by undergoing mergers \citep{white1978}. The Milky Way therefore contains abundant \textit{substructure}, composed of different populations of stars born either inside or outside the Galaxy. Examples of substructure include merged dwarf galaxies such as the Helmi Streams \citep{helmi_streams}, \textit{Gaia} Sausage-Enceladus \citep{belokurov2018,helmi2018}, and Sequoia \citep{myeong2019_sequoia}. These dwarf galaxies are often accompanied by their own populations of globular clusters (GCs), which can be associated with their parent galaxies through a combination of dynamics, chemistry and ages \citep[e.g.][]{MySausageGC,Mas19,kruijssen2020,Cal22}. GCs can in turn produce their own substructure. Stars can escape through the Lagrange points of a cluster (or satellite galaxy) due to the Milky Way's tidal field. The stripped stars often form a stellar stream composed of a pair of tails leading and trailing the cluster approximately along its orbit \citep{binney_tremaine}. To date $\sim100$ stellar streams from GCs or dwarf galaxies have been discovered in the Milky Way \citep{galstreams}.

Interest in GCs and their debris has increased since the arrival of high-redshift data from JWST. Observations suggest that $\sim50\%$ of the stellar mass of some galaxies was contained within massive star clusters at $z\sim8$ \citep{mowla2024}. In the Milky Way, \citet{belokurov2024} divided the population of GCs into accreted and \textit{in situ}. The clusters likely to have been born in the Milky Way are at low energies ($E\lesssim E_\odot$), in the thick disc and inner halo. The low-metallicity ([Fe/H]~$\lesssim-1.3$) \textit{in situ} clusters are associated with the population of stars born in the Milky Way before the formation of the disc at $z\sim3-5$, known as Aurora \citep{belokurov_kravtsov} or the Poor Old Heart \citep{Rix2022}. GCs therefore provide a portal through which to study the earliest stages of our Galaxy's evolution. However, many of these clusters are likely to have been heavily tidally stripped or completely dissolved. JWST observations \citep[e.g.][]{mowla2024} and results from the Milky Way \citep[e.g.][]{martell2010,belokurov2023} suggest that a large fraction of field stars in the halo originated in GCs. Some stars born in GCs can be identified by their unusual chemistry, in particular their high [N/O] ratios \citep[e.g.][]{martell2010,horta2021,schiavon2017,belokurov2023,kane2025}. There is a strong overlap between the distribution of these stars and members of Aurora \citep{kane2025}.

Dynamics are also frequently employed in attempts to identify unbound substructure. Populations accreted or dissolved many Gyr ago are likely to be highly phase-mixed \citep{helmi_white}. This means that there are no longer strong correlations between the orbital phases of stars from the same progenitor. The use of dynamics to uncover substructure therefore relies on integrals of motion (IoMs), which are conserved along each orbit. If the Galactic potential is axisymmetric about the $z$-axis in the region explored by an orbit, the energy $E$ and the $z$-component of angular momentum $L_z$ are exactly conserved \citep{binney_tremaine}. Though the total angular momentum vector precesses about the $z$-axis, its magnitude does not vary significantly. The quantity $L_\perp\equiv\sqrt{L_x^2+L_y^2}$ is therefore often treated as an integral of motion \citep[e.g.][]{helmi_streams}, and is a measure of an orbit's vertical motion above and below the Galactic plane. In an axisymmetric potential any substructure will remain approximately frozen in the 3D space $(L_z,L_\perp,E)$. Any tidally disrupted satellite galaxies or globular clusters should thus remain tightly clustered in this space, even when highly phase-mixed and dispersed across the Galaxy. This is the basis of various attempts to identify substructure from dynamical quantities. These studies typically employ clustering algorithms such as HDBSCAN \citep{hdbscan} in IoM space to identify sets of stars on similar orbits \citep[e.g.][]{lovdal2022,dodd2023,ou2023,liu2024,kim2025}, which may have originated in a common progenitor. An alternative to $(L_z,L_\perp,E)$ is the set of action variables $\textbf{J}$, which (along with the angle variables $\boldsymbol{\theta}$) are related to the Cartesian coordinates (\textbf{x},\textbf{v}) via a canonical transformation \citep[e.g.][]{binney_tremaine}. If they exist, the actions are exactly conserved in an axisymmetric potential and should therefore preserve information about past merger events \citep[e.g.][]{My18}. They are also adiabatic invariants, so remain approximately constant during slow changes to the potential (e.g. gradual mass accretion).

However, the assumption that $(L_z,L_\perp,E)$ and actions are conserved in the Milky Way is not always valid. One reason is that the Galaxy is not axisymmetric; it hosts a bar \citep{Va64,Bl91,Wh92,Bi91,St94,We94} which rotates with a pattern speed (angular frequency) $\Omegab\approx30-45$~km/s/kpc \citep{,Po17,bovy2019,Sa19,binney2020,chiba2021_treering,Cl22,leung2023,zhang24,dillamore2025}, but is likely to be decelerating with time \citep{chiba2021,zhang25}. Along with spiral arms \citep[e.g.][]{sellwood2002}, the bar is capable of changing the angular momenta of stars in the disc and halo \citep[e.g.][]{halle2015,chiba2021,dillamore2023}, resulting in radial migration. It is also well-known that the bar is able to perturb stellar streams. Both the Ophiuchus \citep{Ha16,price-whelan2016,yang2025} and Pal 5 \citep{pearson2017} streams may be perturbed by interactions with the bar, creating fans or gaps, or flipping the tails. \citet{dillamore2024b} showed that a slowing bar can transport globular clusters to higher $L_z$ and $E$ by trapping them in its resonances. Stars stripped from these trapped clusters form diffuse structures instead of narrow streams. The effects of resonances on stream morphology were discussed extensively by \citet{yavetz2021,yavetz2023}.

In any steadily rotating potential such as a barred galaxy, conservation of $E$ and $L_z$ is replaced with conservation of the Jacobi integral $\HJ$ (see Section~\ref{section:theory}). A second integral of motion may only exist in specific scenarios, such as a non-rotating bar or harmonic potential \citep{vandervoort1979}. In general the number of known integrals is therefore reduced to one, though attempts have been made to identify proxies for a second integral \citep{qin2021}. In some cases orbits affected by the bar are chaotic, as recently demonstrated by \citet{woudenberg2025}. Any substructure composed of orbits susceptible to the bar's influence should thus be dispersed in IoM space. This is especially true at low energies, in the regions occupied by Aurora and the debris from the Milky Way's presumed ancient GC population \citep{kane2025}. Understanding how substructures dynamically evolve when influenced by the bar is therefore crucial for connecting current observations to the Milky Way's high-redshift past. In this paper we focus on the effects of the bar on the distribution of substructure in IoM space, particularly phase-mixed structures that cannot be discovered via configuration space alone. This will help to inform future substructure searches about when the bar's influence should be taken into account, and how to mitigate it.

The rest of the paper is arranged as follows. In Section~\ref{section:theory} we present a theoretical model of the dynamical evolution of substructure in a barred galaxy. Sections~\ref{section:test_particles} and \ref{section:auriga} present the methods and results from the test particle and cosmological simulations respectively. Finally we summarise our conclusions in Section~\ref{section:conclusions}.

\section{Theory}\label{section:theory}
\subsection{The Jacobi integral}
In an axisymmetric gravitational potential, both energy $E$ and the $z$-component of angular momentum $L_z$ are conserved integrals of motion. This is no longer true in a rotating non-axisymmetric potential, such as in a galaxy with a bar or spiral arms. If the potential $\Phi$ is rotating with constant pattern speed (angular frequency) $\Omegab$, the Jacobi integral is instead conserved. This is defined as
\begin{equation}\label{eq:jacobi}
    \HJ\equiv E-\Omegab L_z,
\end{equation}
and can be seen as the energy in the frame corotating with the potential \citep[e.g.][]{binney_tremaine}. This is clearer when $\HJ$ is expressed in coordinates ($\mathbf{x}',\mathbf{v}')$ in the non-inertial frame rotating at frequency $\Omegab$ (where $\mathbf{v}'$ is the velocity relative to the rotating frame), 
\begin{equation}
    \HJ=\frac{1}{2}|\mathbf{v}'|^2+\Phi(\mathbf{x}')-\frac{1}{2}\Omegab^2R^2,
\end{equation}
where $R\equiv\sqrt{x'^2+y'^2}$ is the cylindrical radius.

While $\HJ$ is not conserved if $\Omegab$ is time-dependent (e.g. a decelerating bar), it remains a valuable quantity. Consider the evolution of $\HJ$ in a potential $\Phi(\mathbf{x}')$ rotating with time-varying frequency $\bOmegab(t)=\Omegab(t)\mathbf{\hat{z}}$. In the corotating frame the equation of motion is
\begin{equation}
    \frac{\mathrm{d}\mathbf{v}'}{\mathrm{d}t}=-\boldsymbol{\nabla}\Phi-2\bOmegab\times\mathbf{v}'-\bOmegab\times(\bOmegab\times\mathbf{x}')-\frac{\mathrm{d}\bOmegab}{\mathrm{d}t}\times\mathbf{x}',
\end{equation}
where the second, third and fourth terms on the right-hand side are the Coriolis, centrifugal, and Euler forces. The Jacobi integral therefore evolves according to
\begin{align}
    \frac{\mathrm{d}\HJ}{\mathrm{d}t}&=\mathbf{v}'\cdot\frac{\mathrm{d}\mathbf{v}'}{\mathrm{d}t}+\mathbf{v}'\cdot\boldsymbol{\nabla}(\Phi-\frac{1}{2}\Omegab^2R^2)-\Omegab\frac{\mathrm{d}\Omegab}{\mathrm{d}t}R^2\\
    &=-\frac{\mathrm{d}\Omegab}{\mathrm{d}t}R(\mathbf{v}'\cdot\boldsymbol{\hat{\phi}}+R\Omegab)\\
    &=-\frac{\mathrm{d}\Omegab}{\mathrm{d}t}L_z.
\end{align}
Combined with equation~\eqref{eq:jacobi} this implies that the energy and angular momentum evolution of a particle are related by
\begin{equation}\label{eq:E_change}
    \frac{\mathrm{d}E}{\mathrm{d}L_z}=\Omegab(t).
\end{equation}
Hence even in a potential rotating at a variable frequency, particles are instantaneously restricted to move along lines of gradient $\Omegab(t)$ in the $(L_z,E)$ plane. These are contours of constant $\HJ$. 

\subsection{Evolution of substructure}
Now consider a set of particles originating from a small area of $(L_z,E)$ space, such as debris from a dissolved globular cluster or dwarf galaxy \citep[e.g.][]{helmi2000}. If $\Omegab$ is fixed, they can spread out along a fixed line in this space (if they experience perturbations from the bar or spiral arms). With a time-varying $\Omegab$ the gradient of these lines changes with time, so they are able to spread into two dimensions.

We can create a simple toy model for this process by treating it as diffusion in angular momentum, analogous to Brownian motion. We assume that interactions with the bar are random encounters which give each star an angular impulse $\Delta L_z$ drawn from some distribution. This assumption is only valid for substructure away from resonances, for which the impulses from different encounters are uncorrelated and unbiased. In this case each particle will perform a random walk in $L_z$, with a corresponding change in energy at each encounter of $\Delta E=\Omegab \Delta L_z$. The diffusion in $L_z$ is described by a diffusion coefficient,
\begin{equation}\label{eq:diffusion_coefficient}
    D\equiv\frac{1}{2}\frac{\overline{(\Delta L_z)^2}}{\tau},
\end{equation}
where $\overline{(\Delta L_z)^2}$ is the mean squared change in $L_z$ per encounter and $\tau$ is the typical time between encounters with the bar \citep{einstein1905}. Since $D$ depends on the probability distribution of $\Delta L_z$ and the time between encounters, it is a function of the orbit's actions (or energy) and $\Omegab$. However, if we consider a single substructure diffusing across a small region of $(L_z,E)$ space, we can approximate $D$ as being independent of $L_z$ and $E$ (but generally time-dependent). In this case the 2D distribution of particles $P(L_z,E,t)$ will obey the diffusion equation,
\begin{equation}\label{eq:diffusion}
    \frac{\partial P}{\partial t}=D\left(\frac{\partial}{\partial L_z}+\Omegab(t)\frac{\partial}{\partial E}\right)^2P,
\end{equation}
where the derivatives account for the fact that particle motion is instantaneously along lines of gradient $\mathrm{d}E/\mathrm{d}L_z=\Omegab(t)$. We give a derivation of equation~\eqref{eq:diffusion} in Appendix~\ref{section:diffusion}.

The evolution of an initially localised set of stars is given by the Green's function solution to this equation. Fourier transforming equation~\eqref{eq:diffusion} in $L_z$ and $E$ gives
\begin{equation}
    \frac{\partial \tilde{P}}{\partial t}=-D\left(k_{L_z}+\Omegab(t)k_E\right)^2\tilde{P},
\end{equation}
where $\tilde{P}(k_{L_z},k_E,t)$ is the Fourier transform of $P$. This has the solution
\begin{align}
    \tilde{P}(k_{L_z},k_E,t)&=\tilde{P}_0(k_{L_z},k_E)\,\mathrm{exp}\left[-\int_0^t D\left(k_{L_z}+\Omegab(t)k_E\right)^2\mathrm{d}t\right],\\
    \tilde{P}_0(k_{L_z},k_E)&\equiv\mathrm{exp}(-ik_{L_z}L_{z0}-ik_EE_0)
\end{align}
where $(L_{z0},E_0)$ is the initial location of the particles. The evaluated integral will contain averages of $\Omegab$ weighted by the diffusion coefficient. For convenience we define the averaged quantities,
\begin{align}
    \overline{D}&\equiv\frac{1}{t}\int_0^tD\,\mathrm{d}t,\label{eq:D_mean}\\
    \langle\Omegab^n\rangle_D&\equiv\frac{1}{\overline{D}t}\int_0^t\Omegab^n(t)D\,\mathrm{d}t,\label{eq:Omega_b_mean}
\end{align}
so the integral can be evaluated as
\begin{equation}
    \tilde{P}(k_{L_z},k_E,t)=\tilde{P}_0\,\mathrm{exp}\left[-\overline{D}\left(k_{L_z}^2+2\langle\Omegab\rangle_D k_{L_z}k_E+\langle\Omegab^2\rangle_D k_E^2\right)t\right].
\end{equation}
The terms in the exponent can be factorised as
\begin{align}
    [...]&=-\frac{1}{2}\begin{pmatrix}k_{L_z}&\Omega_0 k_E\end{pmatrix}\tilde{\boldsymbol{\Sigma}}^{-1}\begin{pmatrix}k_{L_z}\\\Omega_0 k_E\end{pmatrix},\\
    \tilde{\boldsymbol{\Sigma}}^{-1}&\equiv2\overline{D}t\begin{pmatrix}
        1 & \langle\Omegab\rangle_D/\Omega_0 \\
        \langle\Omegab\rangle_D/\Omega_0 & \langle\Omegab^2\rangle_D/\Omega_0^2
    \end{pmatrix},
\end{align}
where $\Omega_0$ is a fiducial frequency ensuring all vectors and matrices are dimensionally consistent, and $\tilde{\boldsymbol{\Sigma}}$ is the covariance matrix in the dimensionally consistent Fourier space $(k_{L_z},\Omega_0 k_E)$. Inverting the Fourier transform gives the Green's function solution,
\begin{align}
    P(L_z,E,t)&=\frac{1}{2\pi|\boldsymbol{\Sigma}|^{1/2}}\mathrm{exp}\left[-\frac{1}{2}(\mathbf{u}-\mathbf{u}_0)^\mathrm{T}\boldsymbol{\Sigma}^{-1}(\mathbf{u}-\mathbf{u}_0)\right],\\
    \mathbf{u}&\equiv(L_z, E/\Omega_0),\\
    \mathbf{u}_0&\equiv(L_{z0}, E_0/\Omega_0),\\
    \boldsymbol{\Sigma}&=\tilde{\boldsymbol{\Sigma}}^{-1}=2\overline{D}t\begin{pmatrix}
        1 & \langle\Omegab\rangle_D/\Omega_0 \\
        \langle\Omegab\rangle_D/\Omega_0 & \langle\Omegab^2\rangle_D/\Omega_0^2
    \end{pmatrix}.\label{eq:Sigma}
\end{align}
The particles therefore spread out into a Gaussian, described in the scaled $(L_z, E/\Omega_0)$ space by the covariance matrix $\boldsymbol{\Sigma}$. The eigenvalues $\lambda_\pm$ and eigenvectors $\mathbf{e}_\pm$ of $\boldsymbol{\Sigma}$ in this space are
\begin{align}
    \lambda_\pm&=\overline{D}t\left(1+q\pm\sqrt{(1-q)^2+4p^2}\right),\\
    \mathbf{e}_\pm&=\left(1-q\pm\sqrt{(1-q)^2+4p^2},2p\right),\\
    p&\equiv\frac{\langle\Omegab\rangle_D}{\Omega_0},\\
    q&\equiv\frac{\langle\Omegab^2\rangle_D}{\Omega_0^2}.
\end{align}
It is convenient to define
\begin{equation}
    \alpha\equiv\frac{\langle\Omegab^2\rangle_D}{\langle\Omegab\rangle_D^2}-1,
\end{equation}
which is the weighted variance of $\Omegab$ over time in units of its squared weighted average. Hence $\alpha$ is small in realistic scenarios where $\Omegab$ changes by less than an order of magnitude.\footnote{For example, if $D$ is constant and $\Omegab$ decays exponentially by a factor of 2, then $\alpha\approx0.04$.} In this case, to first order
\begin{align}
    \lambda_+&\approx2\overline{D}t\left(1+p^2\right)\left[1+\frac{p^4}{\left(1+p^2\right)^2}\alpha\right],\\
    \lambda_-&\approx2\overline{D}t\frac{p^2}{1+p^2}\alpha.
\end{align}
For small $\alpha$ we see that $\lambda_+\gg\lambda_-$, so the cloud of particles will spread out in a narrow strip along the direction of the eigenvector $\mathbf{e}_+$. This is given approximately by
\begin{equation}
    \mathbf{e}_+\approx\left(1-\frac{p^2}{1+p^2}\alpha,p\right),
\end{equation}
and corresponds to lines in $(L_z,E)$ space with gradient
\begin{align}
    \frac{\mathrm{d}E}{\mathrm{d}L_z}&\approx\left(1+\frac{p^2}{1+p^2}\alpha\right)\langle\Omegab\rangle_D\label{eq:gradient}\\
    &\approx\langle\Omegab\rangle_D.
\end{align}
The gradient is approximately equal to the weighted average of the pattern speed across the time of diffusion, where the weighting is by the diffusion coefficient. The present-day gradient of substructure therefore encodes information about the past evolution of the bar's pattern speed. Any discrepancy between the observed slope of diffused substructure in $(L_z,E)$ space and measurements of the current pattern speed would provide evidence of a time-varying $\Omegab$. If the bar is slowing with time \citep[e.g.][]{chiba2021}, the gradient will exceed the current pattern speed.

\begin{figure*}
  \centering
  \includegraphics[width=\textwidth]{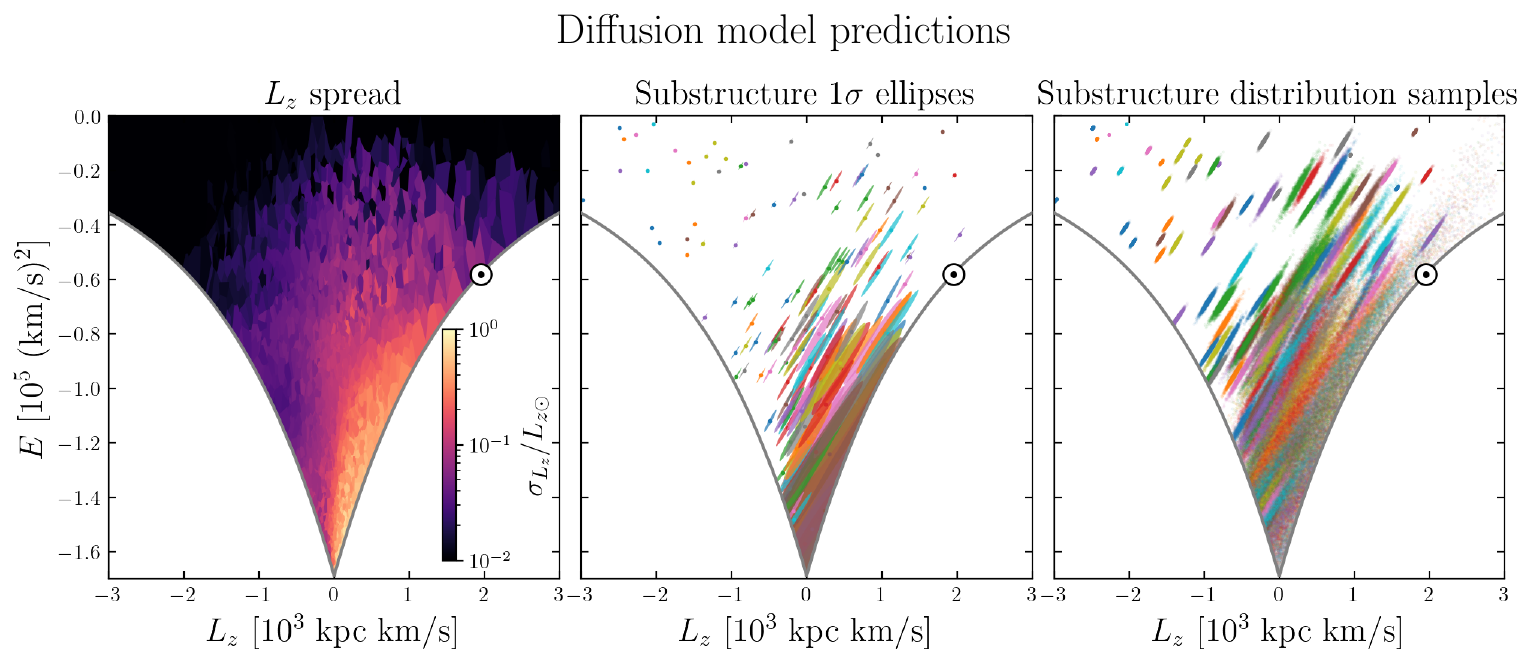}
  \caption{Predictions of the diffusion model. \textbf{Left-hand panel:} predicted angular momentum spread of substructure as a function of $L_z$ and $E$. The $\odot$ symbol indicates a circular orbit at the Sun's radius $r_0$. The bar heavily affects prograde substructure at around the Sun's energy $E_\odot$ and below. \textbf{Middle panel:} $(L_z,E)$ distributions of 150 randomly chosen substructures. The ellipses represent the 1$\sigma$ levels of the Gaussian distributions. The bar causes the substructures to be significantly elongated along one principal direction. \textbf{Right-hand panel:} samples drawn from the same 150 substructures. At $L_z>0$ and $E\lesssim E_\odot$ the substructure is significantly blurred, and localised clumps are not predicted to survive.} 
   \label{fig:diffusion_model}
\end{figure*}

\begin{figure}
  \centering
  \includegraphics[width=\columnwidth]{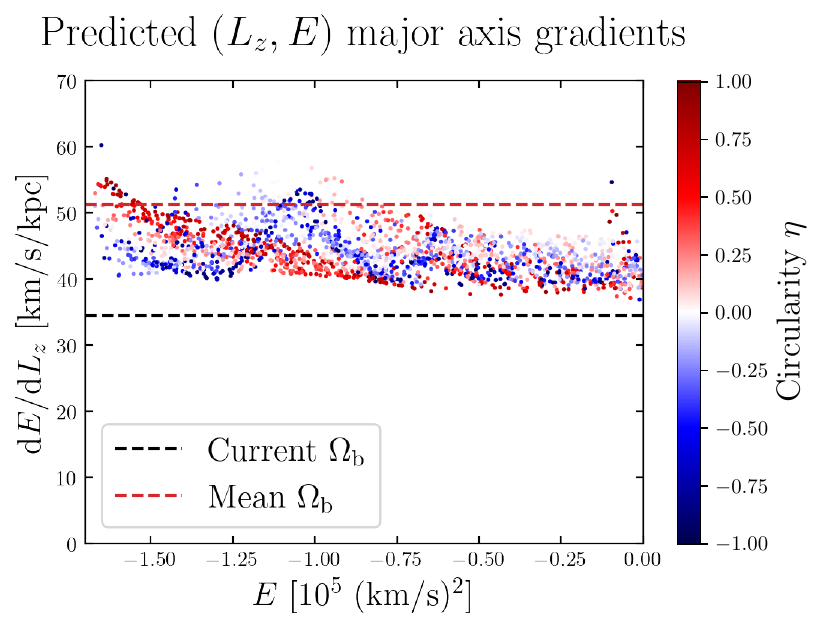}
  \caption{Gradients of the major axes of the substructure Gaussian distributions versus energy. The points are coloured by the circularity of the progenitor orbits. The dashed black and red lines indicate the current and mean pattern speeds $\Omegab$; in most cases the gradient lies between these two values.} 
   \label{fig:predicted_gradients}
\end{figure}
\subsection{Diffusion Coefficient}
According to our model the debris from each substructure will spread into a Gaussian with axis lengths and gradients which depend on the diffusion coefficient $D$. This in turn depends on the geometry of the stars' interactions with the bar. If they are assumed to occur at each pericentric passage, the time between interactions is given by the radial orbital period $\tau\approx2\pi/\Omega_r$, where $\Omega_r$ is the radial frequency. The diffusion coefficient from equation~\eqref{eq:diffusion_coefficient} then becomes
\begin{equation}
    D\approx\frac{\Omega_r\overline{(\Delta L_z)^2}}{4\pi}.
\end{equation}
The change in energy $\Delta E=\Omegab\Delta L_z$ over a radial orbital period is given by
\begin{align}
    \Delta E&=\int_0^\tau\frac{\partial\Phi}{\partial t}(\mathbf{x})\mathrm{d}t\\
    &\approx\frac{1}{\Omega_r}\int_0^{2\pi}\frac{\partial\Phi}{\partial t}(\mathbf{x}_0)\mathrm{d}\theta_r,\label{eq:Delta_E}
\end{align}
where in the second line we have used the impulse approximation to replace the true orbit $\mathbf{x}(t)$ with the unperturbed orbit in the axisymmetric potential $\mathbf{x}_0(t)$, parameterised by the radial angle $\theta_r=\Omega_rt$. 

To proceed we need to assume some form for the potential $\Phi$. We decompose it into a constant spherical potential and a rotating bar component,
\begin{equation}
    \Phi(\mathbf{x},t)=\Phi_0(r)+\delta\Phi(\mathbf{x},t).
\end{equation}
We take $\Phi_0$ to be the isochrone potential used by \citet{dillamore2024},
\begin{align}
    \Phi_0(r)=-\frac{GM}{a+\sqrt{a^2+r^2}},
\end{align}
where $M=2.35\times10^{11}M_\odot$ and $a=3$~kpc. This gives a circular speed at radius $r_0=8.2$~kpc of $v_0=238$~km/s \citep{bland-hawthorn2016,Po17}. While only a moderately good fit to the Milky Way, the isochrone potential has the advantage that all orbits are analytic \citep{binney_tremaine}, so the integrand in equation~\eqref{eq:Delta_E} can be calculated without numerically integrating any orbits.

We use the quadrupole bar potential \citep{chiba2021,hamilton2023},
\begin{align}
    \delta\Phi(\mathbf{x},t)&=\Phi_\mathrm{b}(r)\,\mathrm{sin}^2\theta\,\mathrm{cos}\left[2\left(\phi-\phi_\mathrm{b}\right)\right],\\
    \Phi_\mathrm{b}(r)&=-\frac{Av_0^2}{2}\left(\frac{r}{r_\mathrm{CR}}\right)^2\left(\frac{b+1}{b+r/r_\mathrm{CR}}\right)^5,\\
    \phi_\mathrm{b}&\equiv\int^t\Omegab(t') \mathrm{d}t',
\end{align}
where $A=0.02$ is the bar strength, $r_\mathrm{CR}\equiv v_0/\Omegab$ is an approximation to the corotation radius, and $b=0.28$. This gives the energy change,
\begin{align}
    \Delta E&\approx\frac{2\Omegab}{\Omega_r}\int_0^{2\pi}\Phi_\mathrm{b}(r)\,\mathrm{sin}^2\theta\,\mathrm{sin}\left[2\left(\phi-\phi_\mathrm{b}\right)\right]\,\mathrm{d}\theta_r\\
    &\approx\frac{2\Omegab}{\Omega_r}\int_0^{2\pi}\Phi_\mathrm{b}(r)\,\mathrm{sin}^2\theta\,\mathrm{sin}\left[2\left(\phi-\frac{\Omegab}{\Omega_r}\theta_r\right)\right]\,\mathrm{d}\theta_r,
\end{align}
where in the second line $\Omegab$ is approximated as constant over one orbital period. Without loss of generality $\phi_\mathrm{b}$ has been set to zero at $\theta_r=0$ (i.e. the bar is aligned with $\phi=0$ when the orbit is at pericentre, but the $\phi$ value of the pericentre has not been specified; all orbit orientations are thus included). The coordinates $r$, $\theta$ and $\phi$ are functions of $\theta_r$ along the unperturbed orbit in the potential $\Phi_0$. The relations between these spherical coordinates and the angle variables are given in Section~3.5.2 of \citet{binney_tremaine}. The energy change is a function of the orbit's actions and its orientation. To calculate the diffusion coefficient $D$ as a function of the actions we need to average $(\Delta E)^2$ over all possible orientations for a given set of actions. In the unperturbed potential three angles are required to describe the orientation. The inclination is given by $I=\mathrm{cos}^{-1}(L_z/L)$ and is therefore conserved. However, the longitude of the ascending node relative to the bar's major axis $\Omega$ and the argument of pericentre $\omega$ will both precess. If the substructure is sufficiently phase mixed, these angles can be approximated as being uniformly distributed between 0 and $2\pi$. We therefore average $(\Delta E)^2$ over these two angles. This gives a final expression for the diffusion coefficient as a function of the orbital actions,
\begin{equation}\label{eq:diffusion_coefficient_integral}
    D\!\approx\!\frac{1}{4\pi^3\Omega_r}\!\iint_0^{2\pi}\!\left[\int_0^{2\pi}\!\Phi_\mathrm{b}(r)\,\mathrm{sin}^2\theta\,\mathrm{sin}\!\left[2\left(\phi-\frac{\Omegab}{\Omega_r}\theta_r\right)\right]\!\mathrm{d}\theta_r\right]^2\!\mathrm{d}\omega\mathrm{d}\Omega,
\end{equation}
where the coordinates $r$, $\theta$ and $\phi$ are functions of $\theta_r$ that depend on the actions and the angles $\omega$ and $\Omega$.

We find $D$ at a given set of actions and pattern speed by calculating orbits across a 2D grid of $\omega$ and $\Omega$ values. We use the equations in Appendix~A of \citet{dillamore2024}. We numerically evaluate the $\theta_r$ integral at each gridpoint, then compute $D$ by integrating its square over the grid according to equation~\eqref{eq:diffusion_coefficient_integral}.

\begin{figure*}
  \centering
  \includegraphics[width=\textwidth]{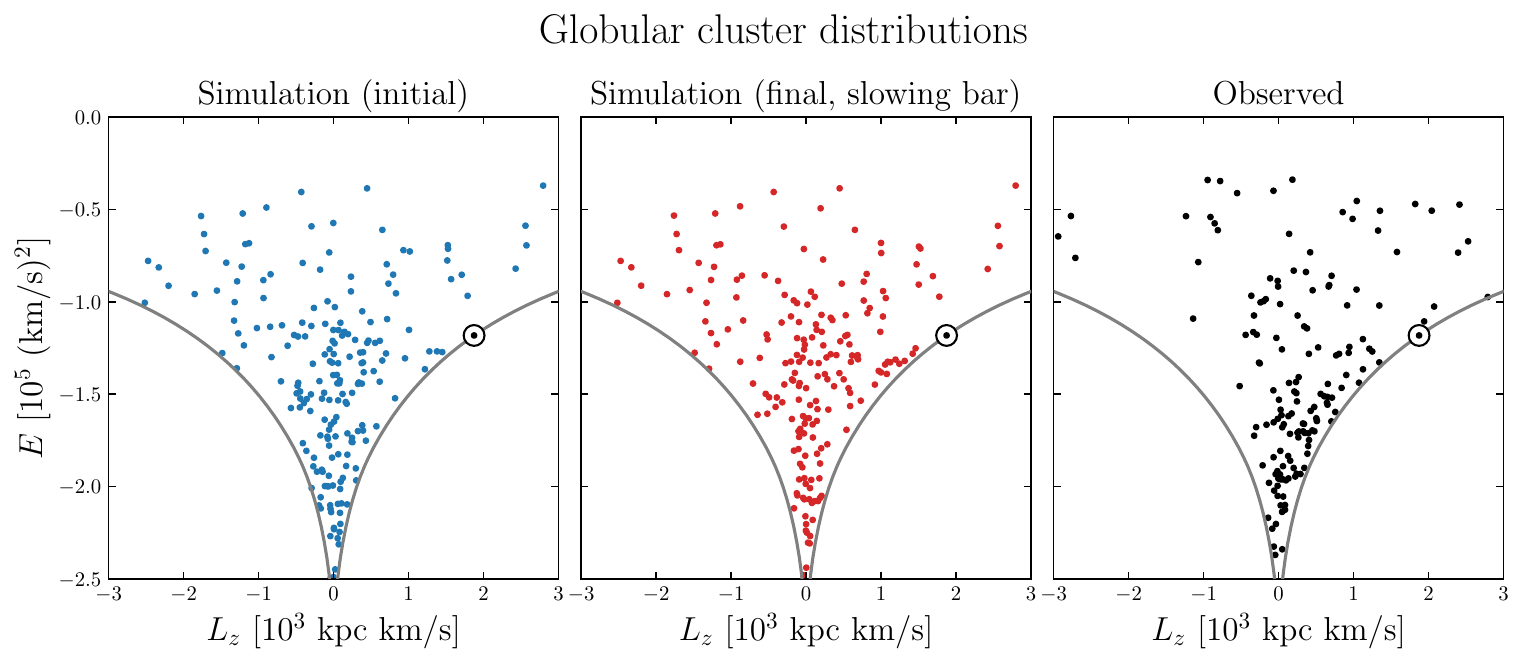}
  \caption{Distributions of globular clusters in $(L_z,E)$ space. \textbf{Left-hand panel:} Initial distribution of mock clusters drawn from a \textsc{DoublePowerLaw} distribution function. \textbf{Middle panel:} final distribution of mock clusters in the simulation with the slowing bar. \textbf{Right-hand panel:} distribution of observed globular clusters in the \citet{vasiliev2021} catalogue.} 
   \label{fig:GC_E_Lz}
\end{figure*}

\subsection{Predicted distribution of tidal debris}\label{section:model_predictions}
In order to predict the distribution of a series of substructures, we must first specify the pattern speed as a function of time. We set
\begin{equation}\label{eq:Omega_b_model}
    \Omegab(t)=\frac{\Omega_\mathrm{b0}}{1+\eta\,\Omega_\mathrm{b0}t},
\end{equation}
where the dimensionless deceleration coefficient $\eta=-\dot{\Omega}_\mathrm{b}/\Omegab^2=0.003$ and $\Omega_\mathrm{b0}=80$~km/s/kpc. This deceleration rate is consistent with \citet{chiba2021} and \citet{zhang25}. We let the period of deceleration be $t\in[0,t_\mathrm{f}]$ where $t_\mathrm{f}\approx5.5$~s~kpc/km~$\approx5.4$~Gyr, consistent with the bar being $\sim8$~Gyr old \citep{sanders2024}. This gives a final pattern speed of $\Omegab(t_\mathrm{f})=34.5$~km/s/kpc, similar to recent estimates for the Milky Way \citep{binney2020,chiba2021_treering,zhang24,dillamore2025}. In the case of a time-independent diffusion coefficient, this pattern speed evolution gives $\alpha\approx0.067$.

We also need to specify the initial distribution function (DF) of substructures in action space. We choose the action-based \textsc{DoublePowerLaw} DF \citep{posti2015,agama},
\begin{align}\label{eq:DF}
    f(\mathbf{J})&\propto\left[1+\left(\frac{J_0}{h(\mathbf{J})}\right)^\eta\right]^{\Gamma/\eta}\left[1+\left(\frac{g(\mathbf{J})}{J_0}\right)^\eta\right]^{-B/\eta},\\
    g(\mathbf{J})&\equiv g_rJ_r+g_zJ_z+(3-g_r-g_z)|J_\phi|,\\
    h(\mathbf{J})&\equiv h_rJ_r+h_zJ_z+(3-h_r-h_z)|J_\phi|.
\end{align}
We take the DF parameters from the fit by \citet{wang2022} to the Milky Way's globular clusters. The outer and inner slopes are $B=5.03$ and $\Gamma=1.23$ respectively, and the transition steepness is $\eta=1.08$. The parameters controlling the flattening and anisotropy are $g_r=0.65$, $g_z=1.32$, $h_r=1.86$, and $h_z=1.01$. Finally the action scale is $J_0=10^{3.08}$~kpc~km/s. We do not include any net rotation, so that we equally sample prograde and retrograde orbits. We draw samples from this distribution using \textsc{agama} \citep{agama}, giving us a mock globular cluster-like distribution of substructure progenitors in action space.

For each progenitor we calculate $D$ over a grid of times $t$ using equation~\eqref{eq:diffusion_coefficient_integral}, then evaluate $\overline{D}$ and $\langle\Omegab^n\rangle_D$ using equations~\eqref{eq:D_mean} and \eqref{eq:Omega_b_mean}. This allows us to calculate the covariance matrices $\boldsymbol{\Sigma}$ of the substructure distributions in $(L_z,E)$ space according to equation~\eqref{eq:Sigma}.

We show the results in Fig.~\ref{fig:diffusion_model}. The left-hand panel shows the predicted angular momentum spread $\sigma_{L_z}=(2\overline{D}t)^{1/2}$ as a function of $L_z$ and $E$. We normalise the values of $\sigma_{L_z}$ by the angular momentum of a circular orbit at the Sun's radius, $L_{z\odot}=r_0v_0\approx2000$~kpc~km/s (indicated by the $\odot$ symbol). The bar is predicted to heavily affect substructure on prograde orbits at lower energies than the Sun. This is because low-energy prograde orbits move slowly relative to the rotating bar, so interactions are stronger and the diffusion coefficient is larger.

To produce the middle and right-hand panels we randomly select 200 mock progenitors from the distribution, roughly equal to the number of globular clusters in the Milky Way \citep[e.g.][]{vasiliev2021}. The middle panel shows the $1\sigma$ ellipses of the predicted Gaussian distributions calculated from equation~\eqref{eq:Sigma}. As expected, the major axes of the ellipses have positive gradients (related to the time dependence of the pattern speed). The dispersion is most significant at $L_z>0$ and $E<E_\odot$, causing the distributions of different clusters to significantly overlap at low energies. Meanwhile at $L_z<-1\times10^3$~kpc~km/s or high energies the amount of spreading is negligible compared to the separation of the clusters, suggesting that substructure can survive in these regions.

In the right-hand panel we draw 1600 samples from each of the 200 Gaussian distributions to represent mock stellar populations. At $L_z>0$ and $E<E_\odot$ the blurring effect is so severe that different clusters would be indistinguishable without the colour-coding. The model predicts that phase mixed substructure can only remain tightly clustered at $E\gtrsim E_\odot$ or $L_z<0$. This challenges the notion that ancient substructure in the inner Milky Way (e.g. dissolved globular clusters) can be discovered using clustering in dynamical spaces alone, unless they are significantly younger than the bar \citep[or the bar is younger than current estimates; e.g.][]{sanders2024}.

However, even where substructure is blurred, the gradients of the distributions' major axes may still be discerned.  We plot the gradients calculated from equation~\eqref{eq:gradient} as a function of energy $E$ in Fig.~\ref{fig:predicted_gradients}. The points are coloured by the circularity $\eta\equiv L_z/L_z(E)$, where $L_z(E)$ is the angular momentum of the circular orbit at energy $E$. Red (blue) points indicate prograde (retrograde) orbits. The black and red dashed lines mark the final and time-averaged pattern speed in our model. A large majority of the points lie between the final and average values of $\Omegab$. This is because the gradient for each substructure is approximately equal to the \emph{weighted} mean of the pattern speed, $\langle{\Omegab}\rangle_D$, where the weighting is by the diffusion coefficient $D$. Since changes to $L_z$ tend to be larger when the pattern speed is slower, the diffusion of stars is faster at lower values of $\Omegab$. The weighting by $D$ therefore decreases the gradients, resulting in most lying between the current and average values of $\Omegab$. Since they preserve memory of the past evolution of the pattern speed, they present a possible method of constraining the time dependence of the Milky Way's bar. If bar-dispersed substructure can be identified in the Milky Way (e.g. by chemistry), the slope can be measured and compared to the current pattern speed inferred by other methods.

Our diffusion model has produced several valuable predictions for the distribution of substructure under the influence of the Galactic bar. We now turn to more realistic numerical models to verify these predictions.

\section{Test particle simulations}\label{section:test_particles}
An assumption of the diffusion model above is that the substructure is already phase-mixed before the bar forms. In reality this may not be true, since stars recently stripped from a cluster will be correlated in orbital angles and hence interactions with the bar. Test particle simulations allow us to drop this assumption by releasing particles from the Lagrange points of a population of clusters. We describe the setup of the simulation below.

\begin{figure*}
  \centering
  \includegraphics[width=\textwidth]{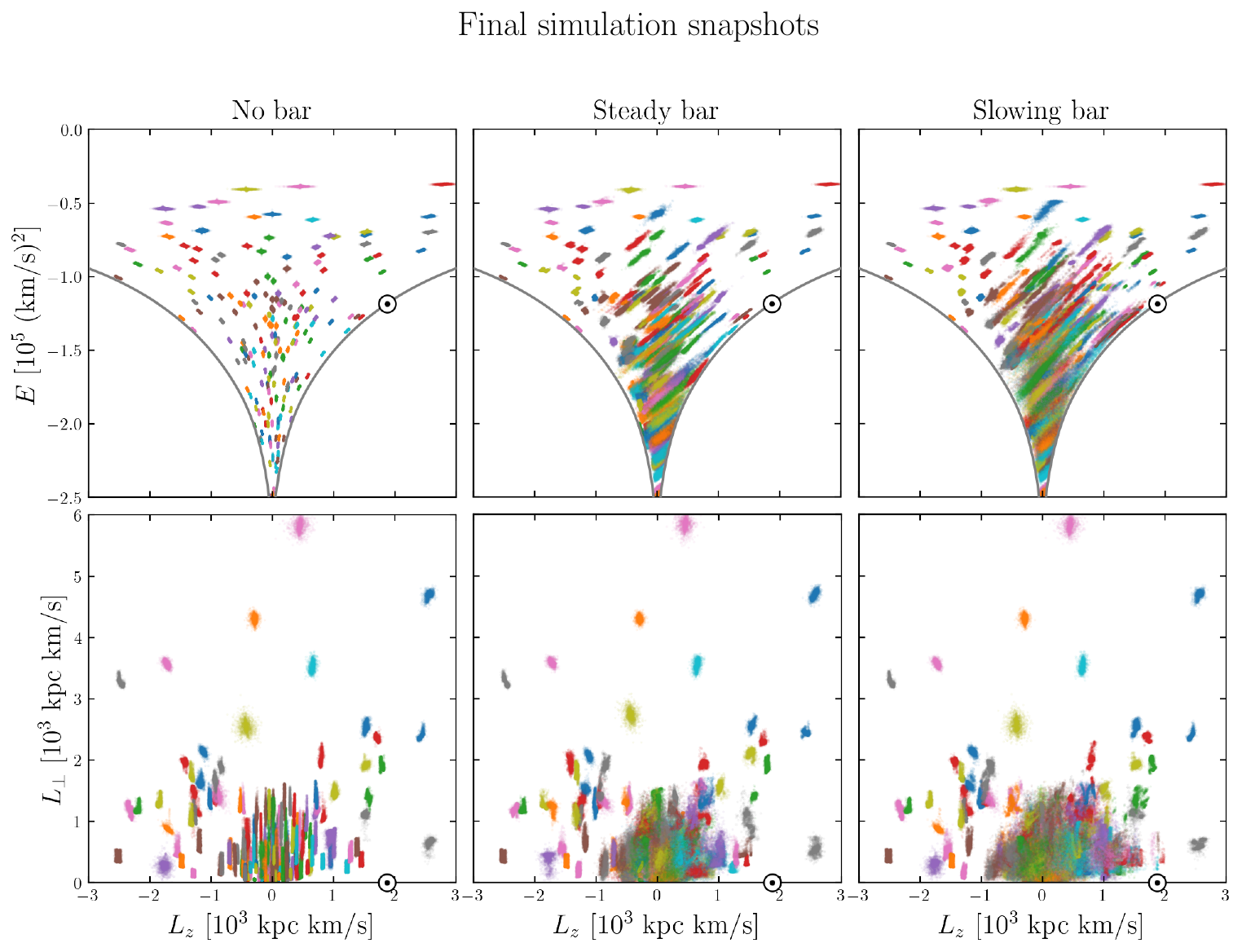}
  \caption{Final snapshots of the test particle simulations in $(L_z,E)$ space (top row) and $(L_z,L_\perp)$ space (bottom row). \textbf{Left-hand column:} The axisymmetric potential, in which $L_z$ and $E$ are conserved for all particles. \textbf{Middle column:} the potential with a steadily rotating bar of pattern speed $\Omegab=34.5$~km/s/kpc, where the Jacobi integral $H_\mathrm{J}$ is conserved. \textbf{Right-hand column:} the potential with a decelerating bar, where the pattern speed decreases from $\Omegab=80$ to $34.5$~km/s/kpc. The $\odot$ symbol marks a circular planar orbit at the Sun's Galactocentric radius.} 
   \label{fig:E_Lz_sim}
\end{figure*}

\subsection{Setup}
We run the simulation in a barred potential similar to that used by \citet{dillamore2025}. This is based on the barred Milky Way potential by \citet{hunter2024}, which includes the bar potential fitted by \citet{sormani2022} to the \citet{Po17} model. We modify the \citet{hunter2024} potential by scaling the non-axisymmetric Fourier components such that the relative length $S(t)=\Omegab(t_\mathrm{f})/\Omegab(t)$, where $\Omegab(t_\mathrm{f})$ is the final pattern speed. This means that the bar length is roughly proportional to the corotation radius. We increase the bar strength smoothly between times $t=1$~s~kpc/km~$\approx1$~Gyr and $t=2$~s~kpc/km according to equation~(4) in \citet{dehnen2000}. The pattern speed is initially $\Omega_\mathrm{b0}=80$~km/s/kpc and begins to smoothly decrease between $t=2$ and $3$~s~kpc/km. It then decreases with constant deceleration parameter $\eta=-\dot{\Omega}_\mathrm{b}/\Omegab^2=0.003$ until $t=t_\mathrm{f}=8$~s~kpc/km, at which $\Omegab(t_\mathrm{f})\approx34.5$~km/s/kpc, similar to our diffusion model. For comparison with the slowing bar, we also run simulations in the axisymmetric (azimuthally averaged) potential, and in the \citet{hunter2024} potential rotating at constant pattern speed $\Omegab(t_\mathrm{f})$.

We generate a mock population of 200 globular clusters in the initial axisymmetric potential using the distribution function given by equation~\eqref{eq:DF}, with the same parameters as in Section~\ref{section:model_predictions}. We integrate their orbits in the steady and slowing bar potentials using \textsc{agama} \citep{agama}. Their initial distribution is plotted in the left-hand panel of Fig.~\ref{fig:GC_E_Lz}, and their final distribution in the simulation with the slowing bar is shown in the middle panel. For comparison the right-hand panel shows the observed Milky Way GCs in the same potential, using the catalogue from \citet{vasiliev2021}.

The main difference between the initial and observed distributions is that our mock population is roughly symmetric in $L_z$, while there are more observed GCs with $L_z>0$. This choice means that we initially equally sample both prograde and retrograde orbits. However, the slowing of the bar affects this symmetry. The middle panel shows that by the end of the simulation, there is an excess of mock GCs at $L_z\sim1\times10^3$~kpc~km/s and $E\sim-1.4\times10^5$~(km/s)$^2$. This is due to trapping and dragging in the bar's corotation resonance, which increases the angular momentum of the population and gives it a net prograde bias. This effect may contribute to the excess of globular clusters on prograde disc-like orbits in the Milky Way \citep{dillamore2024b}.

We generate tidally stripped debris by releasing particles from the two Lagrange points of each cluster \citep[e.g.][]{gibbons2014,bowden2014,fardal2015}. For a cluster at position $\mathbf{x}$, the Lagrange points are approximately located at $\mathbf{x}_\mathrm{t}=\mathbf{x}(1\pm r_t/|\mathbf{x}|)$, where the tidal radius is
\begin{equation}
    r_\mathrm{t}\equiv\left(\frac{GM_\mathrm{c}}{\Omega^2-\frac{\partial^2\Phi}{\partial r^2}}\right)^{1/3}.
\end{equation}
Here $M_\mathrm{c}$ is the mass of the cluster and $\Omega$ is its instantaneous angular speed about the Galactic centre \citep[e.g.][]{bowden2014,gibbons2014}. The particles are assigned a velocity drawn from a Gaussian with velocity dispersion $\sigma$. Following \citet{bowden2014}, the centre of this Gaussian has the same radial velocity as the cluster, while the tangential velocity components match those of the points halfway between the cluster and the Lagrange points. We set $M_\mathrm{c}=1\times10^5M_\odot$, which is a typical globular cluster mass \citep{baumgardt2018}. The velocity dispersion is kept fixed at $\sigma=1$~km/s, consistent with models of globular cluster streams \citep[e.g.][]{bowden2014,dillamore2022b}. We release one particle from each Lagrange point every 10 Myr, giving a total of 1600 stars in the tidal debris of each cluster. The stars' orbits are then evolved in the Galactic potential until the end of the simulation (for $\approx8$ Gyr). To save computation time we do not include the gravity of the progenitors, since we are only interested in the orbits of escaped stars.

We also run a second simulation in each potential with 2000 progenitors sampled from the DF, but with particles released only every 100 Myr. This allows us to measure the overall effects of the bar on substructure with higher resolution in integral of motion space.

\begin{figure}
  \centering
  \includegraphics[width=\columnwidth]{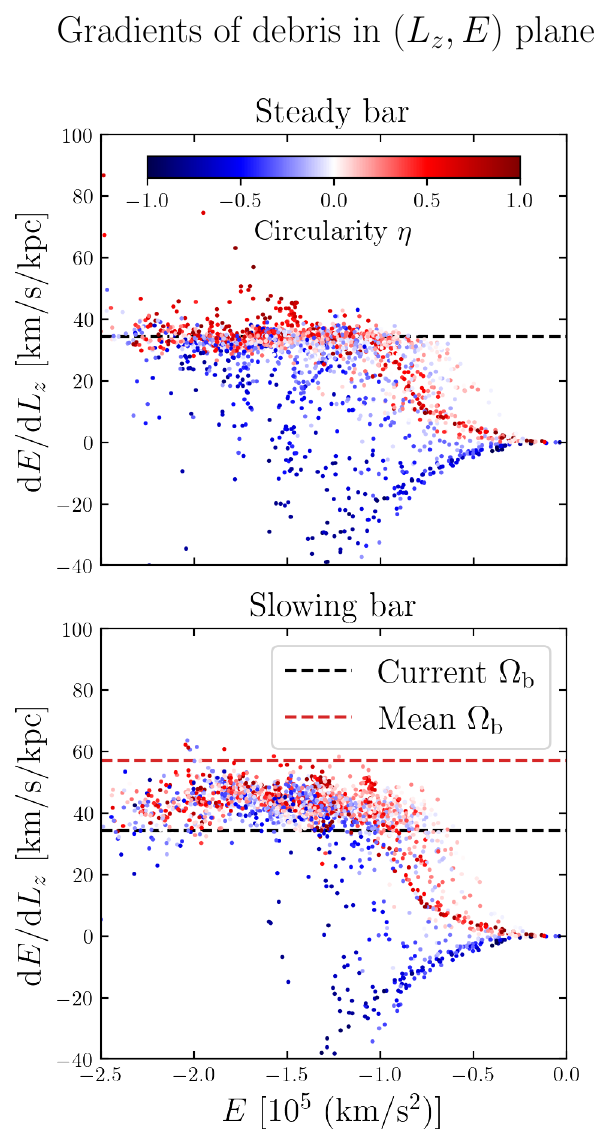}
  \caption{Slopes of the best-fit lines to the debris from each cluster in the final simulation snapshots. The top and middle panels show the simulations with the steady and slowing bars respectively. The black dashed lines indicates the final pattern speed $\Omegab=34.5$~km/s/kpc, while the red dashed line in the lower panel marks the mean $\Omegab$ across the simulation. The points are colour-coded by the clusters' orbital circularity, where red (blue) points correspond to prograde (retrograde) orbits.} 
   \label{fig:gradients_sim}
\end{figure}

\begin{figure*}
  \centering
  \includegraphics[width=0.95\textwidth]{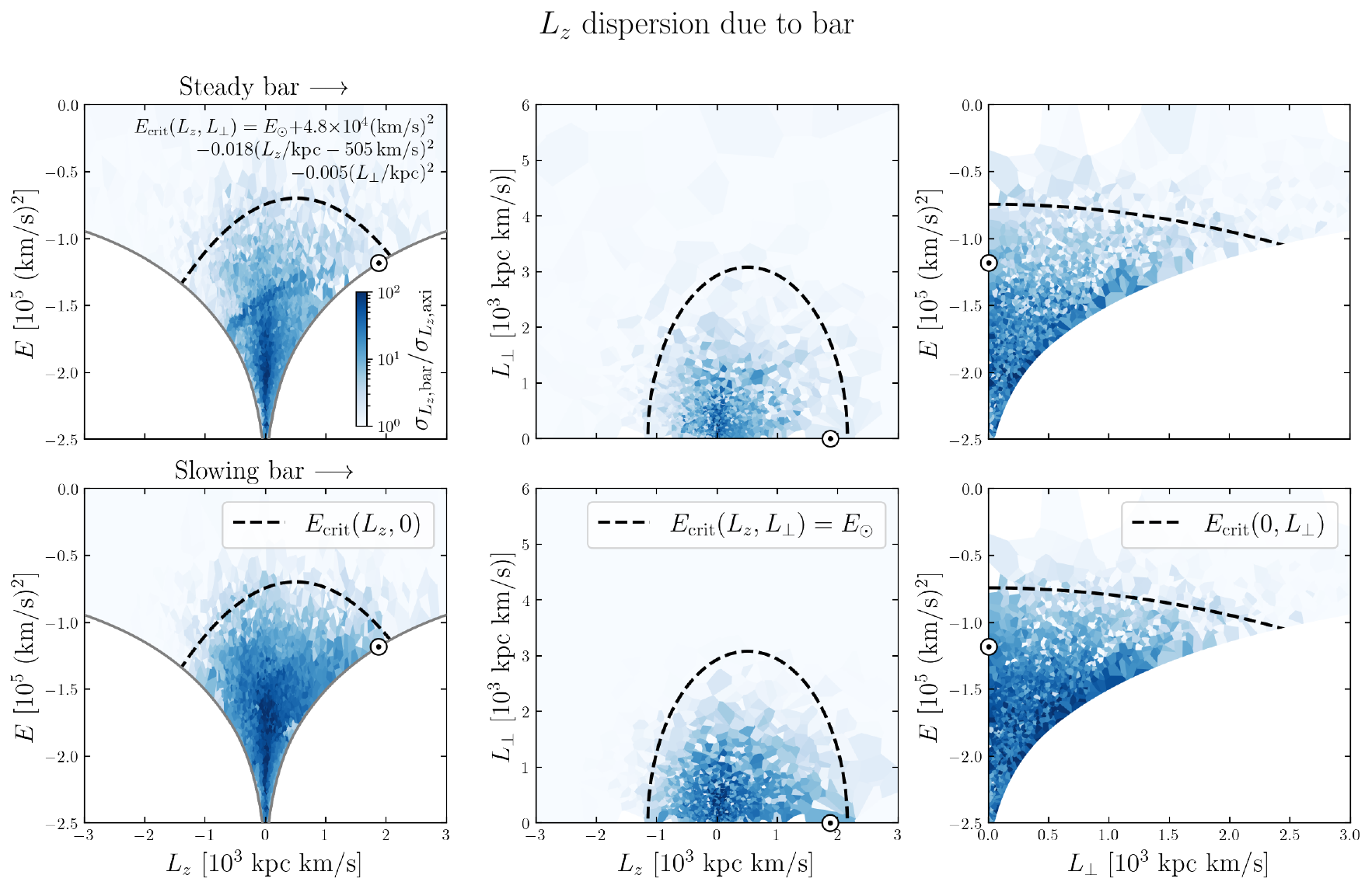}
  \caption{Standard deviation of $L_z$ for each substructure in the barred potentials $\sigma_{L_z,\mathrm{bar}}$ compared to the axisymmetric potential $\sigma_{L_z,\mathrm{axi}}$. The top and bottom rows show the steady and slowing bars respectively. The colour scale runs from white (unaffected by the bar) to dark blue (dipsersed by a factor of 100). The dashed lines show slices of the critical boundary $E_\mathrm{crit}(L_z,L_\perp)$. \textbf{Left-hand column:} $(L_z,E)$ space. The dashed line shows the boundary for orbits in the Galactic plane, $E_\mathrm{crit}(L_z,0)$. \textbf{Middle column:} $(L_z,L_\perp)$ space. In this case the dashed line shows the contour of the boundary at Solar energy, i.e. $E_\mathrm{crit}(L_z,L_\perp)=E_\odot$. \textbf{Right-hand panel:} $(L_\perp,E)$ space. The dashed line marks the boundary for polar orbits, $E_\mathrm{crit}(0,L_\perp)$.} 
   \label{fig:Lz_dispersion_sim}
\end{figure*}

\begin{figure}
  \centering
  \includegraphics[width=\columnwidth]{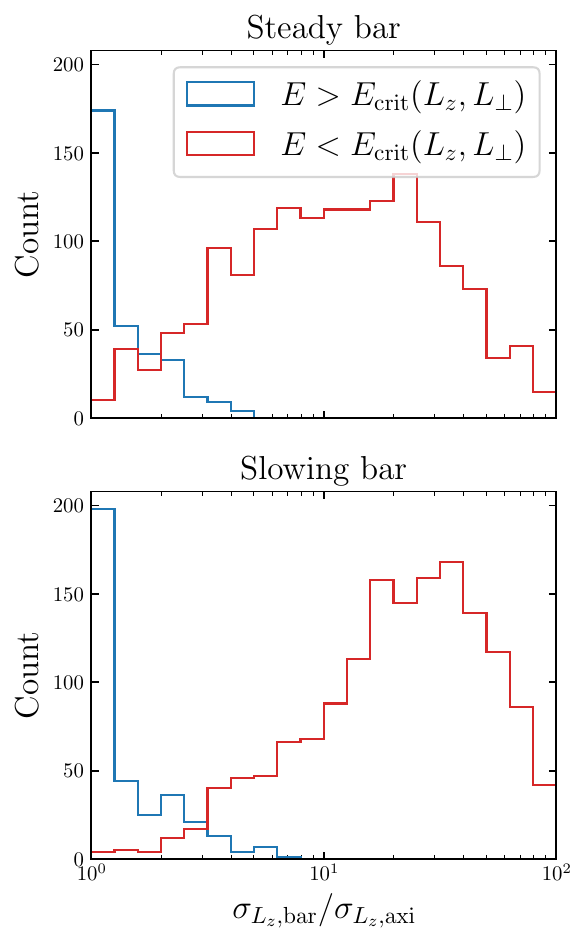}
  \caption{Histograms of the ratio $\sigma_{L_z,\mathrm{bar}}/\sigma_{L_z,\mathrm{axi}}$ for substructure above (blue) and below (red) the critical energy, again for the steady (top panel) and slowing (bottom panel) bars. A large majority of the substructure above (below) the critical energy is dispersed by a factor of $<e$ ($>e$).}
   \label{fig:boundary_test}
\end{figure}

\begin{figure*}
  \centering
  \includegraphics[width=\textwidth]{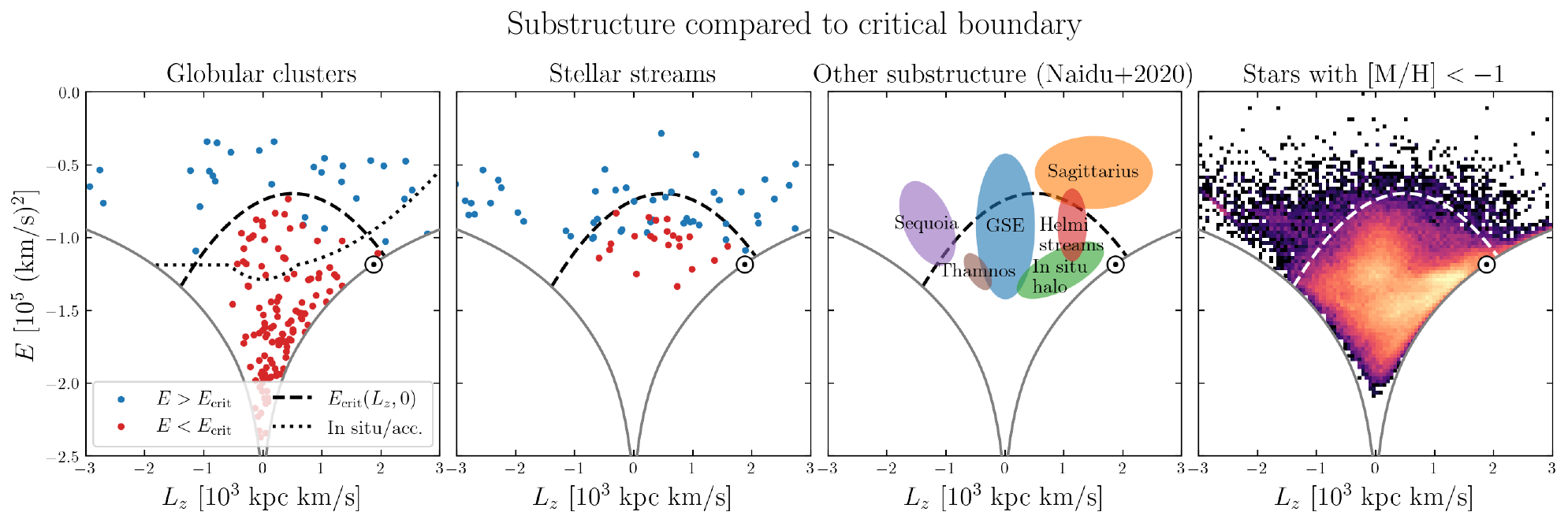}
  \caption{Observed substructure compared to the slice of the critical boundary for orbits in the Galactic plane, $E_\mathrm{crit}(L_z,0)$. \textbf{Left-hand panel}: globular clusters from the \citet{vasiliev2021} catalogue. The points are coloured blue (red) if they lie above (below) the critical boundary in $(L_z,L_\perp,E)$ space. The dotted line marks the approximate boundary between \textit{in situ} (below) and accreted (above) clusters according to \citet{belokurov2024}. \textbf{Second panel:} stellar streams from the orbit fits by \citet{bonaca2025}. A number of streams above the boundary are projected below it in this space due to their high orbital inclinations. \textbf{Third panel:} schematic of selected substructure described by \citet{naidu2020}. \textbf{Right-hand panel:} giant stars with metallicities [M/H]~$<-1$, as estimated by \citet{andrae2023}. In each case many known objects are located below the critical boundary, suggesting they are susceptible to significant dispersion by the bar.}
   \label{fig:substructure_boundary}
\end{figure*}

\begin{figure*}
  \centering
  \includegraphics[width=0.95\textwidth]{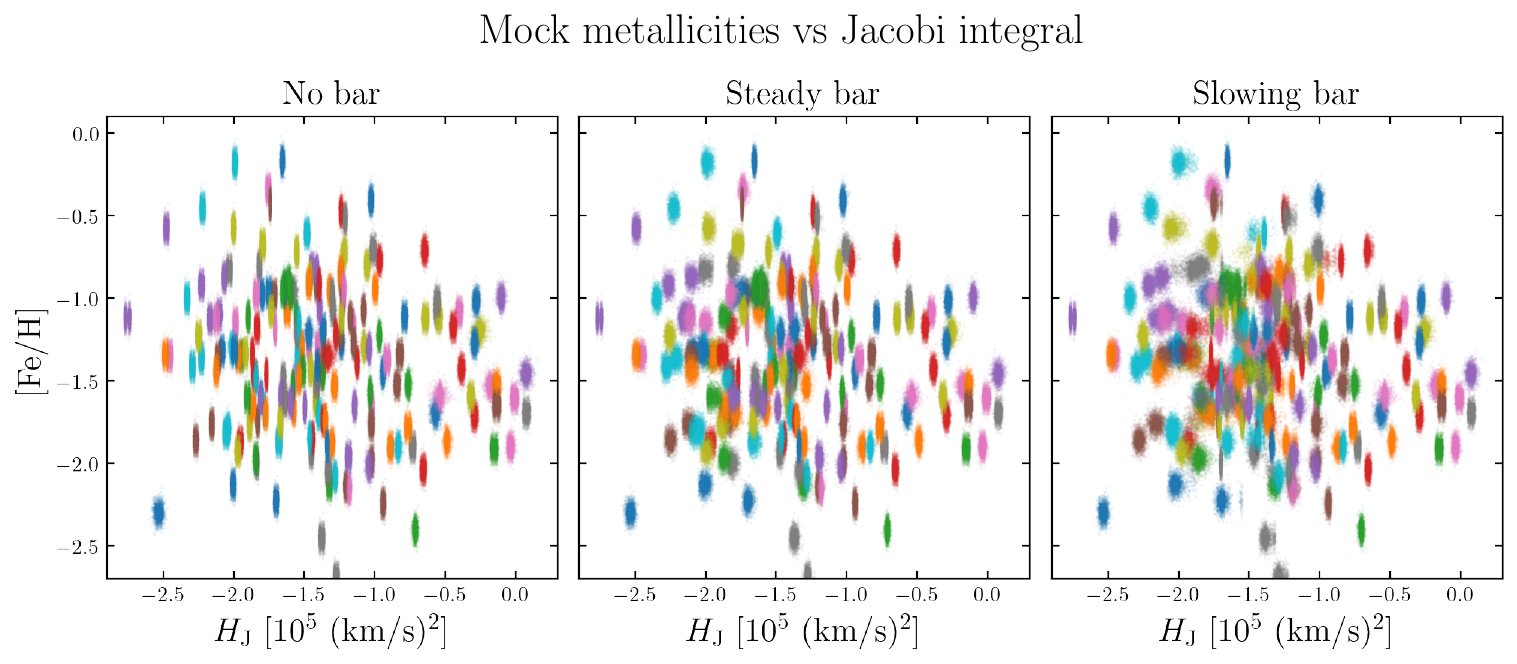}
  \caption{Mock metallicities [Fe/H] vs the Jacobi integral $H_\mathrm{J}$. From left to right, the three panels show the axisymmetric, steady bar and slowing bar potentials. The bar is much less destructive to the clusters in this space than in traditional integral of motion space (see Fig.~\ref{fig:E_Lz_sim}).} 
   \label{fig:FeH_HJ}
\end{figure*}
\subsection{Results}
We show the final $(L_z,E)$ distributions of the debris from all 200 clusters in the top row of Fig.~\ref{fig:E_Lz_sim}. The bottom row shows the $(L_z,L_\perp)$ distributions, where the perpendicular angular momentum $L_\perp\equiv\sqrt{L_x^2+L_y^2}$. The left-hand column shows the simulation in the axisymmetric potential, in which $L_z$ and $E$ are both integrals of motion. The tidal debris therefore remains frozen in this space, and there is no dispersion. The clusters remain similarly compact in $(L_z,L_\perp)$ space. The middle column shows the simulation in the steadily rotating barred potential. In this case the debris from each cluster spreads out significantly from its progenitor. Due to the conservation of the Jacobi integral $H_J$, the movement of each particle in $(L_z,E)$ space is restricted to a fixed line of gradient $\Omegab=34.5$~km/s/kpc. This results in the debris from each cluster spreading out along a narrow strip with positive gradient, as predicted by the diffusion model in Fig.~\ref{fig:diffusion_model}. There is now significant overlap between different individual structures in both $(L_z,E)$ and $(L_z,L_\perp)$ space. The results for the slowing bar in the right-hand panel are similar, though the degree of dispersion of the clusters is even greater. The gradients of the lines traced by each structure in $(L_z,E)$ space are also noticeably steeper. This is consistent with our prediction (equation \eqref{eq:gradient}) that the gradient should equal a weighted average of the pattern speed over time.

We show this effect more clearly in Fig.~\ref{fig:gradients_sim}, where we plot gradients $\mathrm{d}E/\mathrm{d}L_z$ of lines fitted by linear regression to each structure, as a function of progenitor energy. In this case we use the simulations with 2000 progenitors. As in Fig.~\ref{fig:predicted_gradients} we colour-code the points by the orbital circularity $\eta$. The top and bottom panels show the steady and slowing bars respectively. With a steadily rotating bar, most of the gradients are close to the pattern speed (marked by the black dashed line). This is especially true for prograde and low-energy substructure. Retrograde and high-energy substructure tends to have a shallower or even negative gradient. This is because a) the two tails of an unperturbed retrograde stellar stream will be anti-correlated in $(L_z,E)$ space, giving a negative gradient; and b) retrograde structures experience less dispersion due to the bar (see Fig.~\ref{fig:diffusion_model}), so this pre-existing gradient can dominate over any subsequent spreading. The gradient most closely matches the pattern speed when the spreading is large enough to dominate over the initial distribution.

In the simulation with the slowing bar, the gradients at $E\lesssim-1\times10^5$~(km/s)$^2$ almost all exceed the current pattern speed. We also show the (unweighted) mean pattern speed across the full lifetime of the bar with a red dashed line. Nearly all the gradients are less than this mean, in excellent agreement with our predictions from the diffusion model (Fig.~\ref{fig:predicted_gradients}). It demonstrates that the current observable slopes of structures in $(L_z,E)$ space preserve memory of the past evolution of the bar's pattern speed. In particular, Figs.~\ref{fig:predicted_gradients} and \ref{fig:gradients_sim} suggest that the gradients should roughly extend between the current and mean pattern speeds. A series of measured slopes from observations should thus place constraints on the deceleration rate of the Milky Way's bar. This method could complement those using resonances in the Milky Way's disc \citep{chiba2021,zhang25}. At higher energies ($E\gtrsim-1\times10^5$~(km/s)$^2$) there is no correspondence between the gradients and the pattern speed, suggesting that interactions with the bar significantly weaken above this threshold. In Section~\ref{section:crietria} we provide an estimate for the critical boundary below which the bar has a significant effect on the dispersion of substructure.

\subsection{Criterion for substructure survival}\label{section:crietria}
While many structures are significantly blurred by the bar, Fig.~\ref{fig:E_Lz_sim} shows that those at sufficiently high energy are almost unaffected, particularly on retrograde orbits. This is consistent with the predictions of the diffusion model (see Fig.~\ref{fig:diffusion_model}). We now estimate the location of the boundary in $(L_z,E)$ space which separates structures that are significantly affected by the bar from those that are not. We quantify the effect of the bar on a structure by the ratio $\sigma_{L_z,\mathrm{bar}}/\sigma_{L_z,\mathrm{axi}}$, where $\sigma_{L_z,\mathrm{axi}}$ and $\sigma_{L_z,\mathrm{bar}}$ are the standard deviations of $L_z$ of all particles belonging to the structure. This ratio is therefore the factor by which the angular momentum dispersion increases due to the bar. We note that this ratio will increase with the age of the stream for a given progenitor, so it should only be viewed as an order-of-magnitude prediction of disruption.

 We show different projections of phase space colour-coded by this ratio in Fig.~\ref{fig:Lz_dispersion_sim} for the steady (top) and slowing (bottom) bars. Each Voronoi cell corresponds to the initial position of a single progenitor. The colour scale runs from no change (white) to an increase by a factor of 100 (dark blue). From left to right, the three projections are $(L_z,E)$, $(L_z,L_\perp)$, and $(L_\perp,E)$. As expected from the diffusion model and Fig.~\ref{fig:E_Lz_sim}, the dispersion decreases towards higher energies, and prograde orbits are affected more than retrograde orbits. The middle and right-hand panels demonstrate the importance of also taking $L_\perp$ into account. At constant $L_z$ and $E$, the dispersion decreases towards larger $L_\perp$. This is unsurprising; a radial orbit with $L_z=L_\perp=0$ will likely be more affected than a circular polar orbit with $L_z=0$ but large $L_\perp$. To define a critical boundary for disruption by the bar we therefore fit a function $E_\mathrm{crit}(L_z,L_\perp)$ to the simulation data. We choose the simple form of a paraboloid with a peak at $L_\perp=0$ and $L_z>0$,
 \begin{equation}
     E_\mathrm{crit}(L_z,L_\perp)=E_\odot+a-b(L_z/\mathrm{kpc}-c)^2-d(L_\perp/\mathrm{kpc})^2.
 \end{equation}
 We define a structure to be significantly affected by the bar if its dispersion increases by a factor of $\sigma_{L_z,\mathrm{bar}}/\sigma_{L_z,\mathrm{axi}}>e\approx2.7$. We fit the curve to the distribution of initial progenitor coordinates $(L_z,L_\perp,E)_i$ using a weighting to emphasise those with $\sigma_{L_z,\mathrm{bar}}/\sigma_{L_z\mathrm{axi}}\approx e$. Specifically, we use the Gaussian weight,
 \begin{equation}
     w_i=\mathrm{exp}\left(-\frac{\left[\mathrm{ln}(\sigma_{L_z,\mathrm{bar}}/\sigma_{L_z,\mathrm{axi}})_i-1\right]^2}{2\sigma^2}\right),
 \end{equation}
where $\sigma=0.1$. The curve is fitted to all points with $E_i>-2\times10^5$~(km/s)$^2$ using \textsc{scipy.optimize.curve\_fit}, with the uncertainty \textsc{sigma} set to $1/\sqrt{w_i}$. The solution has the equation,
\begin{align}
\begin{split}\label{eq:Ecrit}
    E_\mathrm{crit}(L_z,L_\perp)=E_\odot&+4.8\times10^4\mathrm{(km/s)}^2\\
    &-0.018(L_z/\mathrm{kpc}-505\,\mathrm{km/s})^2\\
    &-0.005(L_\perp/\mathrm{kpc})^2.
\end{split}
\end{align}
Since we include the Solar radius circular orbital energy $E_\odot$ in the equation, it can be used in different Milky Way potential models. The largest apocentre of any orbit with $E<E_\mathrm{crit}$ is $\sim40$~kpc, so equation~\eqref{eq:Ecrit} is approximately valid in any potential similar to \citet{hunter2024} within that radius (though it will change slightly depending on properties such as bar strength and pattern speed). We show slices of this surface in Fig.~\ref{fig:Lz_dispersion_sim} with black dashed lines. From left to right these slices are in the planes $L_\perp=0$, $E_\mathrm{crit}=E_\odot$, and $L_z=0$.

The effectiveness of this boundary in determining the bar's importance is tested in Fig.~\ref{fig:boundary_test}. We divide the structures into two groups according to whether they lie above or below the boundary, and for each simulation plot histograms of the ratio $\sigma_{L_z,\mathrm{bar}}/\sigma_{L_z\mathrm{axi}}$. The blue (red) histograms show structures with energies above (below) $E_\mathrm{crit}(L_z,L_\perp)$. These show that the boundary effectively separates affected and unaffected clusters. With both steady and slowing bars a large majority of the structures above the boundary are dispersed by a factor of less than $\sim e$, and almost all by a factor of less than 5. Hence for structures with $E>E_\mathrm{crit}(L_z,L_\perp)$, it is safe to assume that the extents of their integral of motion space distributions are altered by less than an order of magnitude by the bar. Conversely, a large majority of structures with $E<E_\mathrm{crit}(L_z,L_\perp)$ are dispersed by a factor of more than 2 by both steady and slowing bars. It is therefore probable that any old substructure below the critical threshold has been significantly dispersed by the bar, in many cases by a factor of $>10$.

In Fig.~\ref{fig:substructure_boundary} we compare the boundary $E_\mathrm{crit}(L_z,L_\perp)$ (black dashed line) with observed substructure in the Milky Way. The left-panel shows globular clusters from the \citet{vasiliev2021} catalogue. Those with energy greater than (less than) $E_\mathrm{crit}$ are shown in blue (red). The black dotted line marks the boundary between \textit{in situ} and accreted globular clusters according to \citet{belokurov2024}, where accreted clusters are at higher energy. Over $70\%$ of the clusters lie below $E_\mathrm{crit}(L_z,L_\perp)$, so tidally stripped stars from these clusters are liable to be dispersed far from their progenitors in integral of motion space. This includes almost all \textit{in situ} as well as over 20 accreted clusters.

The second panel shows stellar streams in the catalogue provided by \citet{bonaca2025}. We use the $(L_z,L_\perp,E)$ values of their orbit fits, and shift the energies according to the relative depths of our and their potentials at a radius of 8 kpc. Only $\sim25\%$ of the streams lie below $E_\mathrm{crit}(L_z,L_\perp)$. These include streams previously associated with bar perturbations, in particular Ophiuchus \citep{Ha16,price-whelan2016} and Pal 5 \citep{pearson2017}. We list all streams below the critical boundary in Appendix~\ref{section:affected_streams}. Compared to globular clusters a much smaller fraction of streams are in the region predicted to be affected by the bar, with none at all below $E\sim-1.4\times10^5$~(km/s)$^2$. This may be partly due to selection effects, such as the difficulties of detecting streams in dense stellar fields towards the Galactic centre. However, our results suggest that the total lack of observed streams at low energies may be due to bar-driven dispersal of any globular cluster debris. \citet{baumgardt2019} showed that the initial mass of the Milky Way's globular cluster system was a factor of $\sim5$ times higher than at present, with most of the mass lost from within the Solar orbital radius. We therefore expect that the inner galaxy contains an abundance of undiscovered substructure from tidally stripped globular clusters, much of which is likely to be highly dispersed by the bar.

The third panel of Fig.~\ref{fig:substructure_boundary} shows other known structures in the Milky Way. The plotted ellipses are centred on the median positions reported by \citet{naidu2020}, adjusted for our potential. The ellipse axes and orientations are chosen to roughly match the stellar distributions. Several of these structures are below the critical boundary, including \textit{Gaia} Sausage-Enceladus (GSE) and the \textit{in situ} halo. Compared to dissolved globular clusters, these structures are broader in integral of motion space and contain many more stars. This allows them to be detected even after dispersal by the bar. However, they may have been initially more localised in integral of motion space. High-energy structures such as the Sagittarius stream are not expected to be affected, nor are retrograde structures such as Sequoia \citep{myeong2019_sequoia}. While the Helmi streams are shown below the boundary in this projection, they are on highly inclined orbits \citep[$L_\perp\sim2000$~km/s/kpc;][]{koppelman2019}, so are above the boundary in 3D space. Hence they are not expected to be significantly affected by the bar.

The right-hand panel of Fig.~\ref{fig:substructure_boundary} shows stars from \textit{Gaia} data release 3 \citep[DR3;][]{gaia_dr3}. We use the data from Table 2 of \citet{andrae2023}, which consists of giant stars with metallicities [M/H] estimated from \textit{Gaia} XP spectra. We show stars with parallax signal-to-noise greater than 10 and metallicity [M/H]~$<-1$. This selection includes some accreted stars and \textit{in situ} stars born before the formation of the Milky Way's disc \citep[Aurora/Poor Old Heart][]{belokurov_kravtsov,Rix2022}. A majority of the stars in this sample are below the critical boundary: $\sim95\%$ have $E<E_\mathrm{crit}(L_z,L_\perp)$ (though this is likely to be somewhat enhanced by selection effects). We therefore expect that any unbound substructure in the ancient \textit{in situ} components of the Milky Way will have been significantly dispersed by the bar. This suggests that the high-[N/O] GC-origin stars \citep{belokurov2023,kane2025} are unlikely to remain clustered with others sharing the same progenitor.

\subsection{Effect on clustering}\label{section:clustering}
Searches for small-scale Milky Way substructure often utilise clustering algorithms such as HDBSCAN \citep{hdbscan} in the space of traditional integrals of motion \citep[e.g.][]{lovdal2022,dodd2023,ou2023,liu2024}. The set of quantities $(L_z,L_\perp,E)$ is frequently used. However, we have shown that clusters do not remain confined to small volumes in this space if $E<E_\mathrm{crit}(L_z,L_\perp)$. Fig.~\ref{fig:E_Lz_sim} illustrates that debris from different progenitors is significantly blurred by the bar, causing a large degree of overlap between them. This challenges the notion that clustering algorithms can effectively distinguish debris from different progenitors using the three traditional integrals of motion alone.

We have shown that the spreading of clusters in $(L_z,E)$ space is along narrow streaks of gradient approximately equal to the pattern speed (see Figs.~\ref{fig:E_Lz_sim} and \ref{fig:gradients_sim}). With a steadily rotating bar this is because the Jacobi integral $H_\mathrm{J}$ is exactly conserved. However, even with a slowing bar we can expect debris from a given progenitor to be more tightly clustered in $H_\mathrm{J}$ than in energy or angular momentum. We therefore propose using $H_\mathrm{J}$ instead of $L_z$, $L_\perp$ and $E$ when searching for substructure in the region $E<E_\mathrm{crit}(L_z,L_\perp)$. This can be used in conjunction with chemical abundances such as metallicity [Fe/H]. We illustrate this in Fig.~\ref{fig:FeH_HJ}, where we plot mock metallicity against Jacobi integral for the stars in the simulations (with 200 progenitors). The metallicities are `painted on' to the simulation particles as follows. The mean [Fe/H] for each progenitor is drawn from a Gaussian distribution with mean -1.35 and standard deviation 0.5, which roughly reproduces the metallicity distribution of Milky Way globular clusters \citep{harris1996,harris2010}. For each progenitor the stellar metallicities are drawn from another Gaussian centred on the progenitor's mean [Fe/H], with standard deviation 0.04. This is typical of the metallicity spread of Milky Way globular clusters \citep{latour2025}. We note that the distributions in Fig.~\ref{fig:FeH_HJ} are not entirely realistic. Accreted (\textit{in situ}) globular clusters typically have lower (higher) metallicities and higher (lower) energies \citep{belokurov2024}, so in reality [Fe/H] would anti-correlate with $H_\mathrm{J}$. However, our mock values serve well as a simple illustration.

Fig.~\ref{fig:FeH_HJ} shows that clusters are much less dispersed in $\HJ$ than in $L_z$, $L_\perp$ or $E$ (Fig.~\ref{fig:E_Lz_sim}). The bar only slightly broadens the distributions of $\HJ$, and the clusters remain compact. This is true even in the slowing bar simulation where $\HJ$ is not conserved. Combining $\HJ$ with [Fe/H] allows the clusters to be distinguished and separated across most of the space, and including other chemical abundances may aid this further. We therefore conclude that the Jacobi integral $\HJ$ is a better tool for identifying substructure in integral of motion space than energy or angular momentum separately, at least for $E<E_\mathrm{crit}(L_z,L_\perp)$. This is consistent with the results of \citet{woudenberg2025}.

\begin{figure*}
  \centering
  \includegraphics[width=\textwidth]{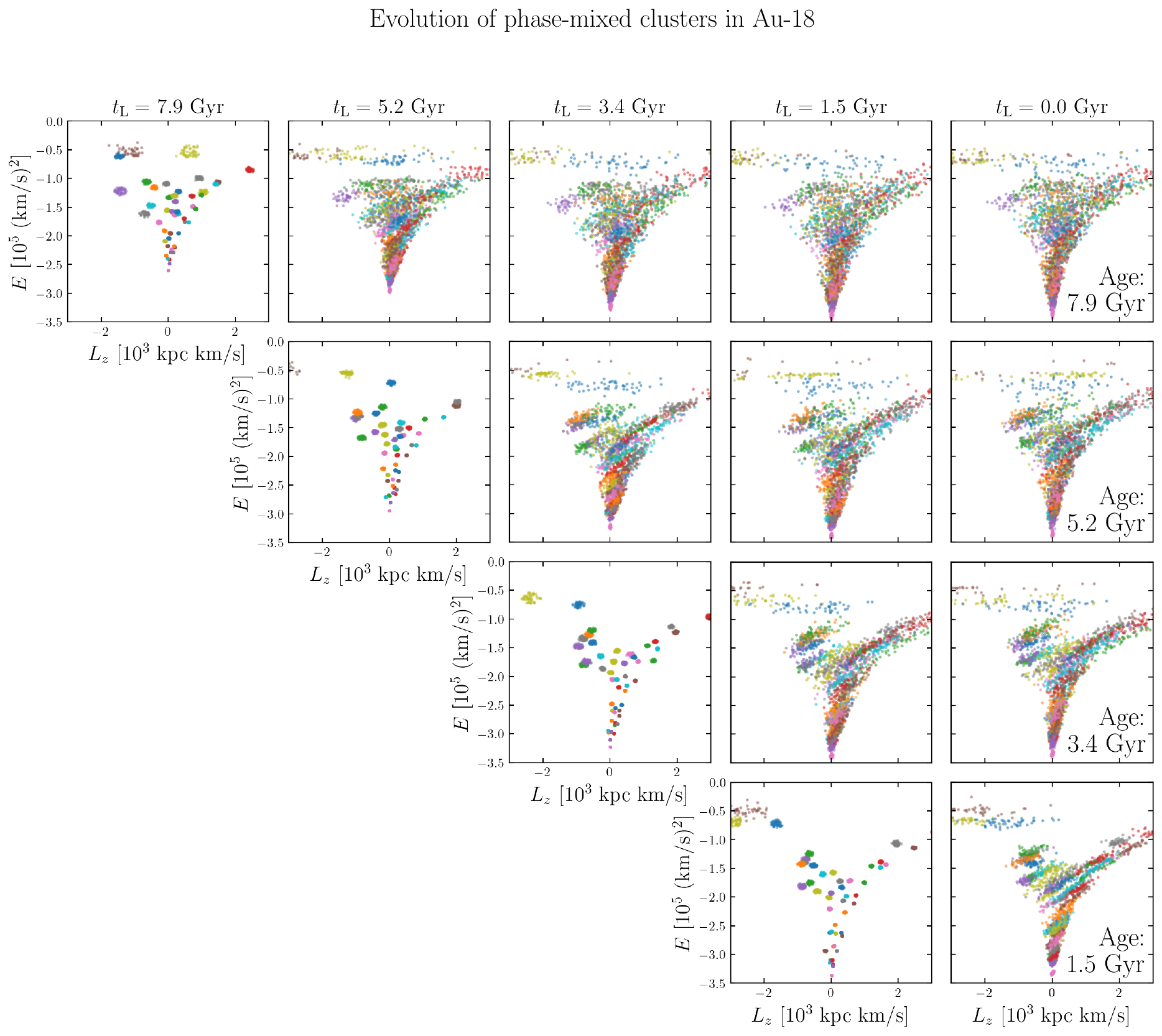}
  \caption{Evolution in $(L_z,E)$ space of mock phase-mixed clusters of stars in Au-18. Each column shows a different snapshot with the lookback time $t_\mathrm{L}$ labelled above. Each row corresponds to sets of stars selected from small regions of integral of motion space at different snapshots. In each row the left-most panel with coloured points is the snapshot at which the particles were selected. From top to bottom, the ages of these mock clusters decrease from 7.9 Gyr to 1.5 Gyr. The results are qualitatively similar to the test particle simulation in Fig.~\ref{fig:E_Lz_sim}; the particles in each cluster spread out along diagonal lines with positive gradient. This is most obvious between the two snapshots in the bottom row.} 
   \label{fig:auriga_E_Lz}
\end{figure*}

\begin{figure*}
  \centering
  \includegraphics[width=\textwidth]{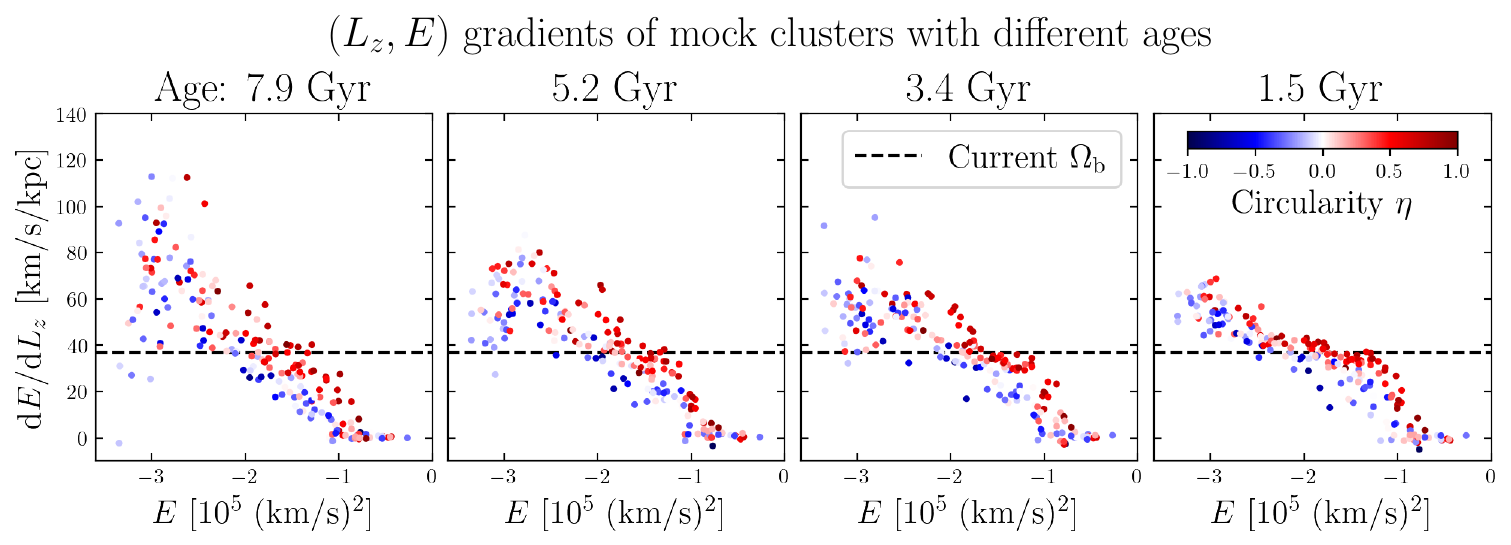}
  \caption{Gradients in $(L_z,E)$ space of different mock clusters at the present day, as in Fig.~\ref{fig:gradients_sim}. Each panel shows a cluster of a different age, with younger towards the right-hand side. As in Figs.~\ref{fig:predicted_gradients} and \ref{fig:gradients_sim}, the colour indicates the orbital circularity and the black dashed line is the final pattern speed. For the younger clusters (right-hand panels), the gradients align closely with the pattern speed, as in the idealised simulation. However, older clusters (left-hand panels) at low energies tend to have steeper gradients.} 
   \label{fig:auriga_gradients}
\end{figure*}

\begin{figure*}
  \centering
  \includegraphics[width=\textwidth]{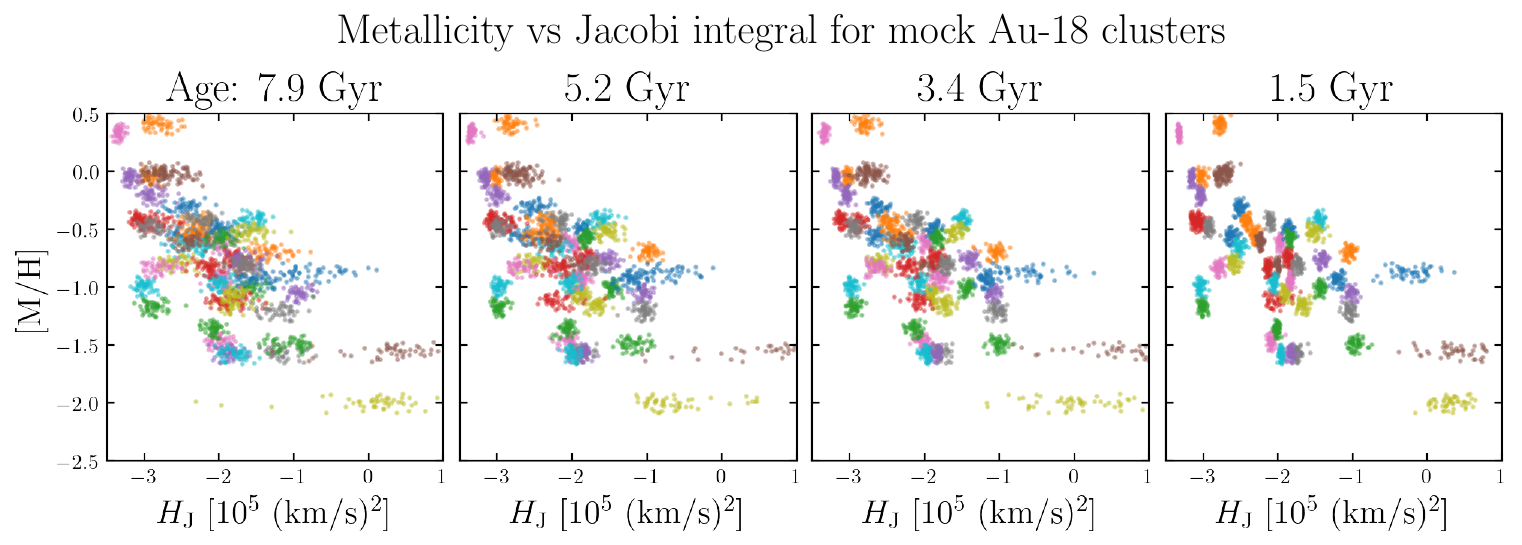}
  \caption{Metallicity vs Jacobi integral at the present day for the mock clusters in Au-18. As in Fig.~\ref{fig:auriga_gradients} the cluster ages decrease from left to right. The clusters are much less dispersed in this space than in $(L_z,E)$ (see Fig.~\ref{fig:auriga_E_Lz}), especially those with younger ages.}
   \label{fig:auriga_MH_HJ}
\end{figure*}

\section{Cosmological simulation}\label{section:auriga}
The test particle simulations described above present an idealised picture of bar perturbations on unbound debris. In reality stars are affected by other perturbations such as spiral arms, minor mergers, galaxy mass growth, and disc reorientation \citep[e.g.][]{dillamore2022}. To take these effects into consideration we now turn to a cosmological zoom-in simulation of a Milky Way-like galaxy from the Auriga suite \citep{grand2017}.

\subsection{The Auriga simulations}
Auriga is a suite of 30 magneto-hydrodynamical zoom-in simulations of galaxies in Milky Way-mass haloes \citep{grand2017}. The haloes have masses in the range $1<M_{200}/(10^{12}M\odot)<2$ and are required to be relatively isolated (i.e. sufficiently far from other massive haloes or galaxy clusters). The simulations are run with cosmological parameters from \citep{planck2014} using the moving mesh code \textsc{arepo} \citep{springel2010}. They include AGN feedback and magnetic fields. We use the level 4 resolution, which has dark matter and baryonic particle masses of $4\times10^5M_\odot$ and $5\times10^4M_\odot$ respectively.

For our study we select the galaxy Auriga 18 (Au-18), which is a close analogue of the Milky Way. It has a Milky Way-like bar \citep{fragkoudi2020} which formed $\sim8$ Gyr before the the present day \citep{merrow2024}, similar to estimates for the Milky Way's bar \citep{sanders2024}. It also has a Milky Way-like merger history, with an analogue of \textit{Gaia} Sausage Enceladus (GSE) merging 9 Gyr ago \citep{fattahi2019}. This produces a comparatively metal-rich component of the halo on highly eccentric orbits, like the Milky Way's GSE \citep{belokurov2018,helmi2018}. It otherwise experiences no major mergers over the last 12 Gyr \citep{fragkoudi2020}. Au-18 is therefore ideal for studying the effects of a bar on substructure in a realistic Milky Way-like environment.

We import the particle data from a set of snapshots of Au-18, and shift the coordinate origin to the galactic centre. We then transform the coordinates to align the $z$-axis with the angular momentum vector of all stars with radii $r<0.05R_{200}$, where $R_{200}$ is the virial radius. In this new coordinate system we compute the angular momentum and energy of each particle, using the potential energy values in the Auriga data. All quantities are converted to physical units. We also calculate the pattern speed of the bar at each snapshot using the method and code provided by \citet{dehnen2023}.

\subsection{Selection of mock clusters}
Since the resolution of the Auriga simulations does not allow stars in globular clusters to be resolved, we cannot trace the evolution of cluster debris without re-simulation. Instead, we construct mock clusters from the star particles in the simulation. In brief, these are sets of stars which share similar values of $L_z$, $L_\perp$ and $E$ at a given snapshot. Since we do not select on orbital phase, the stars in a given cluster are fully phase-mixed from formation. They therefore trace the evolution of dynamically old substructure rather than cold stellar streams.

We first select `progenitor' particles by randomly choosing a set of 200 stars with ages $>12$~Gyr. These can be seen as representing a set of ancient globular clusters, whose ages in the Milky Way mostly exceed 12~Gyr \citep{vandenberg2013}. We now wish to select other star particles with similar angular momenta and energies to represent the stars `stripped' from these progenitors. We therefore define the following distance in integral of motion space,
\begin{equation}
    \Delta I_{ij}^2\equiv(L_{z,i}-L_{z,j})^2+(L_{\perp,i}-L_{\perp,j})^2+(E_i-E_j)^2/\Omega_0^2,
\end{equation}
where $L_{z,i}$, $L_{\perp,i}$, and $E_i$ are the angular momentum components and energy of the $i$th particle, and we take the unit conversion factor $\Omega_0$ to be equal to the pattern speed at the present-day snapshot. We calculate $\Delta I_{ij}$ for each progenitor particle $i$ and each star particle $j$ in each snapshot. The set of 50 particles (including the progenitor itself) with smallest $\Delta I_{ij}$ are taken as the stars stripped from the cluster in that snapshot. These particles are then traced through later snapshots to determine how the phase-mixed cluster evolves in integral of motion space. We also assign metallicities to each cluster, based on the metallicity [M/H] of the progenitor particle. As in Section~\ref{section:clustering}, we assign metallicities to each cluster particle by randomly drawing from a Gaussian centred on the progenitor metallicity with standard deviation 0.04.

\subsection{Results}
The evolution of a set of mock clusters in $(L_z,E)$ space is shown in Fig.~\ref{fig:auriga_E_Lz}. We choose 50 progenitor particles, and select mock clusters around them at four different snapshots, at lookback times $t_\mathrm{L}\in\{7.9,5.2,3.4,1.5\}$~Gyr. The different rows show the evolution of the clusters from each of these times. Each column shows a different snapshot, with the present-day on the right. The results are qualitatively very similar to the predictions of the diffusion model (Fig.~\ref{fig:diffusion_model}) and the test particle simulations (Fig.~\ref{fig:E_Lz_sim}). For each age, the clusters rapidly disperse in $<2$~Gyr. As seen before, this spread is mostly along an axis with a positive gradient. This is most clearly seen between the two snapshots of the 1.5~Gyr-old cluster (bottom row). At higher energies the spread is at approximately constant energy, while $L_z$ is not conserved. This may be explained by a non-axisymmetric potential at large radii. The simulation confirms that in a realistic Milky Way-like galaxy, phase-mixed substructure rapidly disperses and should not be expected to remain tightly bound in integral of motion space.

Like in Fig.~\ref{fig:E_Lz_sim}, we show the gradients of the dispersed substructures at the present day in Fig.~\ref{fig:auriga_gradients}. In this case we select 200 mock progenitors. Each panel shows clusters of a different age, and the colours indicate orbital circularity as before. We show the current pattern speed with the black dashed line. In this simulation $\Omegab$ is relatively steady, varying between $\sim30$~km/s/kpc and $\sim40$~km/s/kpc over its lifetime. The mean pattern speed is therefore close to the current value and is not shown for clarity. In the right-hand panel the results are similar to those for the test particle simulation with the steady bar: below a certain energy most of the structures have gradients close to $\Omegab$. This implies that the bar is driving the spread of these clusters over the last 1.5 Gyr of the simulation. However, the older clusters tend to have steeper slopes at low energies, in many cases greater than the fastest pattern speeds. There must therefore be other processes causing the older clusters to spread out in energy more than angular momentum. This may be due to mass growth of the galaxy. Merger events can cause violent relaxation \citep{lynden-bell1967}, when a rapidly changing potential causes the energies of different particles to vary. This presents an obstacle to using the gradients of dispersed substructure to infer the past pattern speed evolution, since the effects of potential variations may be difficult to disentangle from those of a slowing bar.

The metallicity vs Jacobi integral distributions of 50 clusters are shown in Fig.~\ref{fig:auriga_MH_HJ} for the four different ages. As in the test particle simulation (Fig.~\ref{fig:FeH_HJ}) the clumps are much less dispersed in $\HJ$ than in $(L_z,E)$ space. This is especially true for the younger clusters, suggesting that metallicity/$\HJ$ space may be an effective tool for discovering recently phase-mixed globular cluster debris. However, as the age increases the clusters become more dispersed and overlapping in $\HJ$. This is related to the large spread in gradients seen in Fig.~\ref{fig:auriga_gradients}, and suggests that even $\HJ$ may not be useful for clusters dissolved $\gtrsim10$~Gyr ago.

\section{Conclusions}\label{section:conclusions}
In this paper we have investigated the dynamical effects of the Galactic bar on small-scale substructure, such as dissolved globular clusters. We have employed three different models: an analytic toy model in which the bar's effects are treated as a diffusion process; a set of test particle simulations where stars are stripped from globular clusters and evolved in a realistic Milky Way potential with a slowing bar; and the cosmological zoom-in simulation Auriga 18, which is a close analogue of the Milky Way. Our principal findings are summarised below.

\begin{enumerate}[label=\textbf{(\roman*)}]
    \item The bar causes clusters in $(L_z,E)$ space to disperse along lines of gradient equal to the pattern speed $\Omegab$. If $\Omegab$ is steady, this is equivalent to conservation of the Jacobi integral $\HJ=E-\Omegab L_z$.

    \item We fit straight lines to the distributions of stars from each simulated cluster in $(L_z,E)$ space. With a steadily rotating bar, these gradients are clustered around the pattern speed due to $\HJ$ conservation. With a slowing bar, the present-day gradients are greater than the current $\Omega$ but less than the time-averaged $\Omegab$ across the simulation. The gradients of bar-dispersed substructure in $(L_z,E)$ space therefore preserves memory of the past evolution of the pattern speed. This may offer a route for future studies to constrain the past evolution of the bar's pattern speed. However, the cosmological simulation Au-18 suggests that other processes in real galaxies (e.g. mass growth) may steepen these gradients at low energies, complicating the picture.

    \item We quantify which regions of integral of motion space are most affected by the bar. For each cluster in the test particle simulation we measure the factor by which the bar increases the angular momentum spread $\sigma_{L_z}$ of the stripped debris. We fit a paraboloid surface in $(L_z,L_\perp,E)$ space to the clusters where $\sigma_{L_z}$ increases by a factor of $e$ due to the slowing bar. This surface has the equation,
    \begin{align}
    \begin{split}
        E_\mathrm{crit}(L_z,L_\perp)=E_\odot&+4.8\times10^4\mathrm{(km/s)}^2\\
        &-0.018(L_z/\mathrm{kpc}-505\,\mathrm{km/s})^2\\
        &-0.005(L_\perp/\mathrm{kpc})^2,
    \end{split}
    \end{align}
    where $E_\odot$ is the energy of a circular orbit at the Sun's radius. Any clusters with energy $E<E_\mathrm{crit}(L_z,L_\perp)$ are likely to be significantly affected by the bar. This can be used as a rule of thumb in any realistic Milky Way potential to determine whether the bar is influential. Orbits are more affected at low energy, positive (prograde) $L_z$, high eccentricity, and low inclination.

    \item A significant proportion of the Milky Way's known substructure lies at $E<E_\mathrm{crit}(L_z,L_\perp)$ so is expected to be influenced by the bar. This includes $\sim3/4$ of globular clusters and $\sim1/4$ of known stellar streams (listed in Appendix~\ref{section:affected_streams}). Among these are the Pal 5 and Ophiuchus streams, which have previously been associated with bar perturbations. The lack of observed streams at low energies may be explained by dispersal of globular cluster debris by the bar. Larger-scale substructure such as \textit{Gaia} Sausage-Enceladus and the \textit{in situ} halo are expected to be blurred by the bar. A large majority of the Galaxy's metal-poor ([Fe/H]~$<-1$) stars are also at $E<E_\mathrm{crit}(L_z,L_\perp)$. This includes those born before the formation of the Milky Way's disc \citep{belokurov_kravtsov}. Hence the debris from any globular clusters present in the early Milky Way is expected to be significantly dispersed by the bar. Since much mass has been lost from globular clusters in the inner halo \citep{baumgardt2019}, we predict that it contains a large amount of undiscovered substructure which is no longer clustered in $(L_z,L_\perp,E)$ space. Traditional clustering algorithms are not likely to detect such substructure.

    \item Debris from dissolved clusters is predicted to remain much less dispersed in the Jacobi integral $\HJ$ than in the traditional integrals of motion separately (such as $(L_z,L_\perp,E)$). We show that mock globular cluster debris in the space of metallicity [Fe/H] vs $\HJ$ remains tightly clustered even in a decelerating barred potential. We therefore propose that a combination of chemistry and the Jacobi integral should be used to search for phase-mixed substructure in the inner halo of the Milky Way, particularly in the region $E<E_\mathrm{crit}(L_z,L_\perp)$.
\end{enumerate}
While the bar is expected to have a damaging effect on Milky Way substructure, Galactic archaeology is far from impossible in the inner halo. A combination of the Jacobi integral and chemical abundances should allow different structures to be identified even when highly dispersed in integral of motion space. With the arrival of new chemical abundance measurements from upcoming spectroscopic surveys such as WEAVE, this should bring us a step closer to reconstructing the earliest stages of our Galaxy's evolution.

\section*{Acknowledgements}
We thank the anonymous referee for a helpful report. We are grateful to Vasily Belokurov, GyuChul Myeong, HanYuan Zhang, Sarah Kane, and ChatGPT for helpful discussions and comments during this project. AMD and JLS acknowledge support from the Royal Society (URF\textbackslash R1\textbackslash191555; URF\textbackslash R\textbackslash 241030).

\section*{Data Availability}
This paper uses publicly available \textit{Gaia} data. The code used in this project is available at \url{https://github.com/adllmr/bar_streams}.



\bibliographystyle{mnras}
\bibliography{refs} 




\appendix

\section{Derivation of the diffusion equation}\label{section:diffusion}
Here we derive the diffusion equation~\eqref{eq:diffusion} in the $(L_z,E)$ plane under the approximation of a constant diffusion coefficient $D$.

Consider an infinitesimal bin of area $\mathrm{d}L_z\mathrm{d}E$ at $(L_z,E)$, containing $P(L_z,E,t)\mathrm{d}L_z\mathrm{d}E$ particles at time $t$. In the time interval $t\to t+\Delta t$ particles will experience some change in angular momentum $\pm\Delta L_z$ due to the bar, where we take $\Delta L_z>0$ and assume positive and negative changes are equally likely. Since the energy change is fixed to $\pm\Delta E=\pm\Omegab(t)\Delta L_z$ by equation~\eqref{eq:E_change}, half of the particles experiencing changes $\pm\Delta L_z$ from each of the bins at $(L_z\pm\Delta L_z,E\pm\Omegab(t)\Delta L_z)$ will move to $(L_z,E)$. Meanwhile all of the particles at $(L_z,E)$ experiencing finite $\Delta L_z\neq0$ will be outside the bin at time $t+\Delta t$. Let $f(\Delta L_z)$ be the probability distribution of $\Delta L_z$, assumed to be independent of $L_z$, $E$, and $t$ in the domain of interest. The new occupation of the bin is then obtained by summing over all $\Delta L_z$,
\begin{equation}
\begin{split}
    P(L_z,E,t+\Delta t)=\frac{1}{2}\int_0^\infty[P(L_z-\Delta L_z,E-\Omegab(t)\Delta L_z,t)\\
    +P(L_z+\Delta L_z,E+\Omegab(t)\Delta L_z,t)]\,f(\Delta L_z)\,\mathrm{d}\Delta L_z.
\end{split}
\end{equation}
Expanding both sides and cancelling terms, we obtain
\begin{align}
\begin{split}
    \frac{\partial P}{\partial t}\Delta t&=\frac{1}{2}\left[\frac{\partial^2P}{\partial L_z^2}+2\Omegab(t)\frac{\partial^2P}{\partial L_z\partial E}+\Omegab^2(t)\frac{\partial^2P}{\partial E^2}\right]\\
    &\qquad\times\int_0^\infty(\Delta L_z)^2\,f(\Delta L_z)\,\mathrm{d}\Delta L_z,
\end{split}\\
\begin{split}
    &=\frac{\overline{(\Delta L_z)^2}}{2}\left(\frac{\partial}{\partial L_z}+\Omegab(t)\frac{\partial}{\partial E}\right)^2P.
\end{split}
\end{align}
We can now define the diffusion coefficient,
\begin{equation}
    D\equiv\frac{1}{2}\frac{\overline{(\Delta L_z)^2}}{\Delta t},
\end{equation}
where we let $\Delta t=\tau$ be the average time between interactions with the bar, and $\pm\Delta L_z$ the change in angular momentum per interaction. We can now recover the diffusion equation~\eqref{eq:diffusion},
\begin{equation}
    \frac{\partial P}{\partial t}=D\left(\frac{\partial}{\partial L_z}+\Omegab(t)\frac{\partial}{\partial E}\right)^2P.
\end{equation}

\section{Streams predicted to be affected by the bar}\label{section:affected_streams}
In Table~\ref{tab:affected_streams} we list the streams in the \citet{bonaca2025} catalogue whose energies $E$ are below our critical energy for bar-driven dispersal, $E_\mathrm{crit}(L_z,L_\perp)$. The right-hand column shows the magnitude of the energy difference between the streams and the boundary, shown in descending order. The streams nearer the top of the list are therefore most likely to be heavily perturbed by the bar.

\begin{table}
    \centering
    \begin{tabular}{cc}
        \hline
        Stream & $|E-E_\mathrm{crit}(L_z,L_\perp)|$ [$10^5$ (km/s)$^2$] \\
        \hline
        New-15 & 0.63 \\
        New-6 & 0.48 \\
        New-9 & 0.47 \\
        Gaia-7 & 0.30 \\
        Hydrus & 0.23 \\
        NGC288 & 0.20 \\
        NGC6397 & 0.20 \\
        NGC1851 & 0.16 \\
        Svol & 0.16 \\
        C-9 & 0.16 \\
        New-27 & 0.14 \\
        NGC7099 & 0.13 \\
        Pal5 & 0.11 \\
        M5 & 0.10 \\
        Ophiuchus & 0.08 \\
        Gaia-8 & 0.06 \\
        New-17 & 0.05 \\
        New-19 & 0.04 \\
        LMS-1 & 0.04 \\
        M92 & 0.03 \\
        Hrid & 0.01 \\
        \hline
    \end{tabular}
    \caption{Streams predicted to be affected by the bar. The right-hand column shows the distance in energy below the critical boundary for bar-driven dispersal.}
    \label{tab:affected_streams}
\end{table}


\bsp	
\label{lastpage}
\end{document}